\documentclass[%
 aip,
 amsmath,amssymb,
preprint,%
 fleqn,
]{revtex4-1}

\usepackage{graphicx}
\usepackage{dcolumn}
\usepackage{bm}

\begin{document}


\title{Dust ion acoustic solitary structures at the acoustic speed in presence of nonthermal electrons and isothermal positrons}

\author{Ashesh Paul}
\affiliation{ Department of Mathematics, Jadavpur University, Kolkata
- 700032, India.}%
\author{Anup Bandyopadhyay}
\email{abandyopadhyay1965@gmail.com}%
\affiliation{ Department of Mathematics, Jadavpur University, Kolkata
- 700032, India.}%
\author{K. P. Das}
\affiliation{ Department of Applied Mathematics, University of Calcutta, 92 Acharya Prafulla Chandra Road, Kolkata - 700009, India.}%

\begin{abstract}
\noindent The Sagdeev pseudo-potential technique and the analytic theory developed by Das \textit{ et al.} [\textit{J. Plasma Phys.} \textbf{78}, 565 (2012)] have been used to investigate the dust ion acoustic solitary structures at the acoustic speed in a collisionless unmagnetized  dusty plasma consisting of negatively charged static dust grains, adiabatic warm ions, nonthermal electrons and isothermal positrons. The present system supports both positive and negative potential solitary waves at the acoustic speed, but the system does not support the coexistence of solitary structures of opposite polarity at the acoustic speed. The system also supports negative potential double layer at the acoustic speed, but does not support positive potential double layer. Although the system supports positive potential supersoliton at the supersonic speed, but there does not exist supersoliton of any polarity at the acoustic speed. Solitary structures have been investigated with the help of compositional parameter spaces and the phase portraits of the dynamical system describing the nonlinear behaviour of the dust ion acoustic waves at the acoustic speed. For the case, when there is no positron in the system, there exist negative potential double layer and negative potential supersoliton at the acoustic speed and for such case, the mechanism of transition of supersoliton to soliton after the formation of double layer at the acoustic speed has been discussed with the help of phase portraits. The differences between the solitary structures at the acoustic speed and the solitary structures at the supersonic speed have been analysed with the help of phase portraits. 
\end{abstract}

\maketitle

\section{Introduction}

The investigations of dust ion acoustic (DIA) solitary structures in four component electron-positron-ion-dust (e-p-i-d) plasmas have received a great deal of attention in the last few years as e-p-i-d plasma may be found in numerous cosmic sites such as around the pulsars \cite{shukla04}, near the surface of the neutron stars \cite{zeldovich1971,shukla04}, in the hot spots on dust ring in the galactic centre \cite{zurek1985}, interstellar medium \cite{zurek1985,higdon09,shukla2008}, interior regions of accretion disks near neutron stars and magnetars \cite{dubinov12}, in Milky way \cite{shukla2008}, in the magnetosphere and in the ionosphere of the Earth \cite{alfven1981,gusev2000,gusev2001}, in the magnetosphere of the Jupiter \cite{merlino2006} and the Saturn \cite{horanyi2004} as well as in laboratory environments \cite{shukla04,dubinov12}. Using the reductive perturbation method Ghosh and Bharuthram \cite{ghosh08} investigated the nonlinear propagation of small but finite amplitude ion acoustic (IA) solitons and double layers in a collisionless unmagnetized e-p-i-d  plasma consisting of cold ions, negatively charged static dust particulates and Boltzmann distributed electrons and positrons. Using Bernoulli's pseudo-potential method, Dubinov \textit{et al.} \cite{dubinov12} elaborated the nonlinear theory of DIA waves in a collisionless unmagnetized four component e-p-i-d plasma consisting of warm ions, negatively charged static dust impurities, isothermal electrons and positrons. Several authors \cite{el-tantawy11a,el-tantawy11b,saini13,banerjee16} investigated small or arbitrary amplitude DIA solitary structures in different e-p-i-d plasma systems. Paul \textit{et al.} \cite{paul17ppr} investigated the existence of different DIA solitary structures in a collisionless unmagnetized four component e-p-i-d plasma consisting of negatively charged static dust grains, adiabatic warm ions, isothermally distributed electrons and positrons. They reported the existence of solitary waves of both polarities, coexistence of solitary waves of both polarities, existence of double layers of both polarities, and the existence of positive potential solitons after the formation of positive potential double layer. The existence of positive potential solitons after the formation of double layer confirms the existence of positive potential supersolitons.  Again, Paul \& Bandyopadhyay \cite{paul2016} considered the e-p-i-d plasma system of Paul \textit{et al.} \cite{paul17ppr}, but they considered the Cairns \cite{cairns95} distributed nonthermal electrons instead of isothermal electrons. In this paper, Paul \& Bandyopadhyay \cite{paul2016} extensively discussed the DIA solitary structures with the help of the qualitatively different compositional parameter spaces showing the nature of existence of different solitary structures giving a special emphasis on the existence of solitary structures after the formation of double layer of same polarity. Recently, Paul \textit{et al.} \cite{paul17pop} rigorously studied the formation of supersoliton with the help of the phase portraits of the dynamical system describing the nonlinear behaviour of the DIA waves in a four component e-p-i-d plasma consisting of nonthermal electrons and nonthermal positrons. They clearly discussed the transition process of different solitary structures viz., soliton $\to$ double layer $\to$  supersoliton $\to$ soliton after the formation of double layer  for increasing values of Mach number.

In the above mentioned works, DIA solitary structures have been considered at the supersonic speed only, i.e., for $U>C_{D}$, where $U$ is the velocity of the wave frame and $C_{D}$ is the linearized velocity of the DIA wave for long wave length plane wave perturbation. However, the numerical observations \cite{baluku10,baluku10a,verheest10} of the solitary structures at the acoustic speed, i.e., for $U=C_{D}$, influnced Das \textit{et al.} \cite{das12mc} to set up a general analytical theory for the existence of the solitary structures at the acoustic speed, i.e., for $U=C_{D} \Leftrightarrow M=M_{c}$, where $M=U/C_{D}$ and $M_{c}$ is the lower bound of the Mach number for the existence of solitary structures, i.e., the solitary structures start to exist for $M>M_{c}$. In fact, Das \textit{et al.} \cite{das12mc} have proved three important results to confirm the existence of solitary structures at the acoustic speed. Das \textit{et al.} \cite{das12mc} investigated dust acoustic (DA) solitary structures at the acoustic speed with the help of analytical theory developed in the same paper and they also prescribed a computational scheme to investigate the nature of existence of solitary structures at the acoustic speed. Later, Das \textit{et al.} \cite{das2011existence} investigated DIA solitary structures at the acoustic speed in a collisionless unmagnetized dusty plasma consisting of negatively charged static dust grains, adiabatic warm ions and Cairns distributed nonthermal electrons. They found that the system supports the negative potential solitary waves (NPSWs), positive potential solitary waves,  negative potential double layers (NPDLs) and negative potential supersolitons at the acoustic speed. They also showed the qualitatively different existence domains of DIA solitary structures at the acoustic speed. Recently, Verheest and Hellberg \cite{verheest2015} investigated the existence of IA and DIA solitary structures at the acoustic speed and found the existence of NPDL and negative potential supersoliton at the acoustic speed.

In the present work, following the analytic theory and the computational scheme as developed by Das \textit{et al.} \cite{das12mc}, we have studied the DIA solitary structures at the acoustic speed with the help of the existence domains and the phase portraits of the dynamical system describing the nonlinear behaviour of the DIA waves in the same plasma system considered by Paul \& Bandyopadhyay \cite{paul2016}. In fact, the present paper is an extension of the published work of Paul \& Bandyopadhyay \cite{paul2016} in the following directions.\\
(i) Instead of considering DIA solitary structures at the supersonic speed ($U>C_{D} \Leftrightarrow M>M_{c}$), we have considered different DIA solitary structures at the acoustic speed ($U=C_{D} \Leftrightarrow M=M_{c}$) in a collisionless unmagnetized  four component e-p-i-d plasma consisting of negatively charged static dust grains, adiabatic warm ions, Cairns \cite{cairns95} distributed nonthermal electrons, and isothermal positrons.\\
(ii) For the first time, we have introduced the phase portrait analysis of the dynamical system corresponding to the solitary structures at the acoustic speed ($U=C_{D} \Leftrightarrow M=M_{c}$).\\
(iii) Phase portraits of the dynamical system corresponding to different DIA solitary structures clearly indicate the difference between the different DIA solitary structures at the acoustic speed ($U=C_{D} \Leftrightarrow M=M_{c}$) and the solitary structures at the supersonic speed ($U>C_{D} \Leftrightarrow M>M_{c}$).\\ 
(iv) We have also considered the case when there is no positron in the system and for this particular case, the system supports NPSWs after the formation of NPDL. The existence of NPSWs after the formation of NPDL confirms the existence of negative potential supersolitons. Here we have also discussed the transition process of negative potential solitary structures at the acoustic speed, viz., soliton $\to$ double layer $\to$ supersoliton $\to$ soliton after the formation of double layer . We have seen that the transition process of different solitary structures at the acoustic speed ($U=C_{D} \Leftrightarrow M=M_{c}$) is same as the transition mechanism of solitary structures at the supersonic speed ($U>C_{D} \Leftrightarrow M>M_{c}$).    


\section{\label{sec:basic_eqn}Basic Equations \& Energy Integral}
We consider the exactly same plasma system of Paul \& Bandyopadhyay \cite{paul2016} and consequently we consider the same set of basic equations of Paul \& Bandyopadhyay \cite{paul2016} to study the nature of existence of DIA solitary structures at the acoustic speed. The following are the governing equations describing the nonlinear behaviour of DIA waves propagating along $x$-axis in a collisionless unmagnetized multicomponent dusty plasma system consisting of adiabatic warm ions, negatively charged static dust particulates, nonthermally distributed electrons and isothermal positrons.
\begin{eqnarray}\label{continuity}
\frac{\partial n_{i}}{\partial t}+\frac{\partial}{\partial x}(n_{i}u_{i})=0,
\end{eqnarray}
\begin{eqnarray}\label{momentum}
M_{s}^{2}\bigg(\frac{\partial u_{i}}{\partial t}+u_{i}\frac{\partial u_{i}}{\partial x}\bigg)+\frac{(1-p)\sigma_{ie}}{n_{i}}\frac{\partial p_{i}}{\partial x}+\frac{\partial \phi}{\partial x}=0,
\end{eqnarray}
\begin{eqnarray}\label{pressure}
\frac{\partial p_{i}}{\partial t}+u_{i}\frac{\partial p_{i}}{\partial x}+\gamma p_{i} \frac{\partial u_{i}}{\partial x}=0,
\end{eqnarray}
\begin{eqnarray}\label{poisson}
\frac{\partial^{2} \phi}{\partial x^{2}}=-\frac{M_{s}^{2}-\gamma \sigma_{ie}}{1-p}\bigg(n_{i}-n_{e}+n_{p}-\frac{Z_{d}n_{d0}}{n_{0}}\bigg).
\end{eqnarray}
Here $n_{i}$, $n_{e}$, $n_{p}$, $u_{i}$, $p_{i}$, $\phi$, $x $ and $ t $ are, respectively, the number density of ions, the number density of electrons, the number density of positrons, velocity of ion fluid, ion fluid pressure, electrostatic potential, spatial variable and time, and these have been normalized by $n_{0}$ ($=n_{i0}+n_{p0}=n_{e0}+Z_{d}n_{d0}$), $n_{0}$, $n_{0}$, $C_{D}$ (linearized velocity of the DIA wave in the present plasma system for long-wavelength plane wave perturbation), $n_{i0}K_{B}T_{i}$, $\Phi=\frac{K_{B}T_{e}}{e}$, $ \lambda_{D} $ (Debye length of the present plasma system) and $\lambda_{D}/C_{D}$ with $n_{e0}$, $n_{i0}$, $n_{p0}$ and $n_{d0}$ are, respectively, the equilibrium number densities of electrons, ions, positrons and dust particulates, $ \gamma(=3) $ is the adiabatic index, $ Z_{d} $ is the number of electrons residing on a dust grain surface, $-e$ is the charge of an electron, $T_{i}$ ($T_{e}$) is the average temperature of ions (electrons) and $K_{B}$ is the Boltzmann constant. The expressions of $M_{s}$ and the four basic parameters $p$, $\mu$, $\sigma_{ie}$, $\sigma_{pe}$ are given by the following equations: 
\begin{eqnarray}\label{Ms}
M_{s}=\sqrt{\gamma\sigma_{ie}+\frac{(1-p)\sigma_{pe}}{p+\mu (1-\beta_{e}) \sigma_{pe}}},	
\end{eqnarray}
\begin{eqnarray}\label{p}
p=\frac{n_{p0}}{n_{0}},~\mu=\frac{n_{e0}}{n_{0}},~\sigma_{ie}=\frac{T_{i}}{T_{e}},~\sigma_{pe}=\frac{T_{p}}{T_{e}},	
\end{eqnarray}
where $T_{p}$ is the average temperature of positrons and $\beta_{e}$ is the nonthermal parameter associated with the Cairns model \cite{cairns95} for electron species, and according to  Verheest \& Pillay \cite{verheest08}, the physically admissible bounds of $\beta_{e}$ is given by $0 \leq \beta_{e} \leq \frac{4}{7} \approx 0.6$.

The normalized number densities of nonthermal electrons and isothermal positrons are given by
\begin{eqnarray}\label{ne}
n_{e} = \mu(1-\beta_{e}\phi+\beta_{e}\phi^{2})e^{\phi},
\end{eqnarray}
\begin{eqnarray}\label{np}
n_{p} = p e^{-\phi / \sigma_{pe}}.
\end{eqnarray}
The above equations are supplemented by the following unperturbed charge neutrality condition:
\begin{eqnarray}
n_{i0}+n_{p0}=n_{e0}+Z_{d}n_{d0}.
\end{eqnarray} 

To  investigate the steady state arbitrary amplitude DIA solitary structures, we make all the dependent variables depend only on a single variable $ \xi=x-Mt $, where $M$ is the dimensionless velocity of the wave frame normalized by the linearized DIA speed ($C_{D}$) for long-wavelength plane wave perturbation. Using this transformation and applying the boundary conditions:\\ $ \big(n_{i},p_{i},u_{i},\phi,\frac{d\phi}{d\xi}\big)\rightarrow \big(1-p,1,0,0,0\big)\mbox{    as    }  |\xi|\rightarrow \infty,
$\\ we get the following energy integral:
\begin{eqnarray}\label{energy_integral}
\frac{1}{2}\bigg(\frac{d\phi}{d\xi}\bigg)^{2}+V(\phi)=0,
\end{eqnarray}
where
\begin{eqnarray}\label{V_phi_1}
V(\phi) = (M_{s}^{2}-3\sigma_{ie}) \Big[~V_{i} +\frac{p }{1-p} \sigma_{pe} V_{p}-\frac{\mu}{1-p}V_{e}-\frac{1-\mu}{1-p}V_{d}\Big],
\end{eqnarray}
\begin{eqnarray}\label{V_i_1}
V_{i} = M^{2}M_{s}^{2}+\sigma_{ie} -N_{i}\Big[M^{2}M_{s}^{2}+3\sigma_{ie}-2\phi -2\sigma_{ie}N_{i}^{2}\Big],
\end{eqnarray}
\begin{eqnarray}\label{N_i_1}
N_{i}=\frac{n_{i}}{1-p}=\frac{MM_{s}\sqrt{2}}{(\sqrt{\Phi_{M}-\phi}+\sqrt{\Psi_{M}-\phi})},
\end{eqnarray}
\begin{eqnarray}\label{Phi_M_1}
\Phi_{M} &=& \frac{1}{2}\Big(MM_{s}+\sqrt{3\sigma_{ie}}\Big)^{2}, \\
 \Psi_{M} &=& \frac{1}{2}\Big(MM_{s}-\sqrt{3\sigma_{ie}}\Big)^{2},
\end{eqnarray}
\begin{eqnarray}\label{V_e_1}
V_{e} &=& \big(1+3\beta_{e}-3\beta_{e}\phi+\beta_{e}\phi^{2}\big)e^{\phi}-(1+3\beta_{e}), \\
V_{p} &=& 1-e^{-\phi/\sigma_{pe}}, ~V_{d}=\phi.
\end{eqnarray}
The energy integral (\ref{energy_integral}) can be regarded as the one-dimensional motion of a particle of unit mass whose
position is $\phi$ at time $\xi$ with velocity $d\phi/d\xi$ under the action of the force field $-V'(\phi)$. The first term in the energy integral (\ref{energy_integral}) can be regarded as the kinetic energy of a particle of unit mass at position $\phi$ and time $\xi$, whereas $V(\phi)$ is the potential energy of the same particle at that instant. Now, according to Sagdeev \cite{sagdeev66}, for the existence of a positive (negative) potential solitary wave [PPSW] ([NPSW]) solution of (\ref{energy_integral}), we must have the following three conditions: 
(i) $\phi=0$ is the position of unstable equilibrium of a particle of unit mass associated with the energy integral (\ref{energy_integral}), i.e., $V(0)=V'(0)=0$ and $V''(0)<0$.
(ii) $V(\phi_{m}) = 0$, $V'(\phi_{m}) > 0$ ($V'(\phi_{m}) < 0$) for some $\phi_{m} > 0$ ($\phi_{m} < 0$). This condition is responsible for the oscillation of the particle within the interval $\min\{0,\phi_{m}\}<\phi<\max\{0,\phi_{m}\}$.
(iii) $V(\phi) < 0$ for all $0 <\phi < \phi_{m}$ ($\phi_{m} < \phi < 0$). This condition is necessary to define the energy integral (\ref{energy_integral}) within the interval $\min\{0,\phi_{m}\}<\phi<\max\{0,\phi_{m}\}$. For the existence of a positive (negative) potential double layer [PPDL] ([NPDL]) solution of (\ref{energy_integral}), the second condition is replaced by $V(\phi_{m}) = 0$, $V'(\phi_{m}) = 0$, $V''(\phi_{m}) < 0$ for some $\phi_{m} > 0$ ($\phi_{m} < 0)$. This condition states that the particle cannot be reflected back from the point $\phi=\phi_{m}$ to the point $\phi = 0$.

Therefore, the necessary condition for the existence of solitary waves and / or double layers of any polarity states that $\phi=0$ is the position of unstable equilibrium of a particle of unit mass associated with the energy integral (\ref{energy_integral}), i.e., $V''(0)<0$ along with $V(0)=0$ and $V'(0)=0$. In other words, $\phi=0$ can be made an unstable position of equilibrium if the potential energy of the said particle attains its maximum value at $\phi=0$. Now, from the condition $V''(0)<0$ we get, $M>M_{c}=1$, i.e., the solitary structures (solitary waves and / or double layers) start to exist just above the curve $M = M_{c}=1$. The condition $V''(0)>0$ gives $M<M_{c}=1$. If $M<M_{c}$, the potential energy of the said particle attains its minimum value at $\phi=0$, and consequently $\phi=0$ is the position of stable equilibrium of the particle. In this case, it is impossible to make any oscillation of the particle even when it is slightly displaced from its position of stable equilibrium. Therefore, there is no question of existence of solitary waves or double layers of any polarity for $M<M_{c}$.

Now, let us consider the case for which $V''(0)=0 \Leftrightarrow M=M_{c} \Leftrightarrow V''(M_{c},0)=0$. If $V''(M_{c},0)=0$ along with $V'''(M_{c},0)=0$ then $\phi=0$ is a stable or unstable position of equilibrium according to whether $V''''(M_{c},0)>0$ or $V''''(M_{c},0)<0$. If $V''''(M_{c},0)<0$ then solitary structures may exist at $M=M_{c}$ if the other conditions for the existence of solitary structures are fulfilled. If $V''''(M_{c},0)>0$ then there is no question of existence of solitary structures at $M=M_{c}$. But if $V'''(M_{c},0) \neq 0$ along with $V(M_{c},0)= V'(M_{c},0) = V''(M_{c},0) = 0$, then following the analytic theory developed by Das \textit{et al.} \cite{das12mc}, one can easily study the existence of solitary wave and / or double layer solutions of the energy integral (\ref{energy_integral}) at $M = M_{c}$.

If $V(M,0)=V'(M,0)=V''(M_{c},0)=0$, $V'''(M_{c},0) < 0$ ($V'''(M_{c},0) > 0$), $\partial V/\partial M<0$  for all $M  > 0$ and for all $\phi > 0$ ($\phi < 0$), Das \textit{et al.} \cite{das12mc} have proved the following important results to confirm the existence of solitary structures at the acoustic speed.\\
\textbf{Result-1:} If there exists at least one value $M_{0}$  of $M$ such that the system supports PPSWs (NPSWs) for all $M_{c}< M < M_{0}$, then there exist either a PPSW (NPSW) or a PPDL (NPDL) at $M=M_{c}$.\\
\textbf{Result-2:} If the system supports only NPSWs (PPSWs) for $M>M_{c}$, then there does not exist PPSW (NPSW) at $M=M_{c}$.\\
\textbf{Result-3:} It is not possible to have coexistence of both positive and negative potential solitary structures at $M=M_{c}$.

Again, according to Das \textit{et al.} \cite{das12mc}, the PPDL (NPDL) solution at $M=M_{c}$ is possible only when there exists a PPDL (NPDL)
solution in any right neighborhood of $M_{c}$, i.e., PPDL (NPDL) solution at $M=M_{c}$ is possible only when the curve $M=M_{PPDL}$ ($M=M_{NPDL}$)  tends to intersect the curve $M=M_{c}$ at some point in the existence domain of the energy integral, where each point of the curve $M=M_{PPDL}$ ($M=M_{NPDL}$) corresponds to a PPDL (NPDL) solution of the energy integral whenever $M_{PPDL} > M_{c}$ ($M_{NPDL} > M_{c}$).

From \textbf{Result-1}, \textbf{Result-2} and \textbf{Result-3}, we see that the existence of solitary structures at $M=M_{c}$ depends on the existence of the solitary structures for $M>M_{c}$. Therefore, in the next section, we shall consider qualitatively different existence domains of solitary structures for $M>M_{c}$ to investigate the existence and polarity of the solitary structures at $M=M_{c}$.

\section{\label{sec:solution_spaces} Different existence domains}

From the discussions as given in the previous section, we see that we must have a definite idea regarding the existence and the polarity of the solitary structures in the right neighbourhood of $M=M_{c}$ to apply \textbf{Result-1}, \textbf{Result-2} and \textbf{Result-3} of Das \textit{et al.} \cite{das12mc}  Again, differentiating $V$ with respect to $M$, we get the following equation.
\begin{eqnarray}\label{diff_V_wrt_M}
\frac{\partial V}{\partial M}=-\Bigg\{\sqrt{\frac{M_{s}^{2}M(1-p)\sigma_{pe}}{p+\mu (1-\beta_{e}) \sigma_{pe}}}\Bigg(\sqrt{N_{i}}-\frac{1}{\sqrt{N_{i}}}\Bigg)\Bigg\}^{2}.	
\end{eqnarray}
From equation (\ref{diff_V_wrt_M}), it is simple to check that the following condition holds good.
\begin{eqnarray}\label{condition_1}
\frac{\partial V}{\partial M}<0~~\mbox{for all}~~M>0.	
\end{eqnarray}
Therefore, all the conditions of \textbf{Result-1}, \textbf{Result-2} and \textbf{Result-3} hold good if $V'''(M_{c},0)\neq 0$. So, to discuss the existence and polarity of the solitary structures at $M=M_{c}$, it is necessary to study the qualitatively different existence domains for $M>M_{c}$. It is also necessary to determine the sign of $V'''(M_{c},0)$.
 
\begin{figure}
  \includegraphics{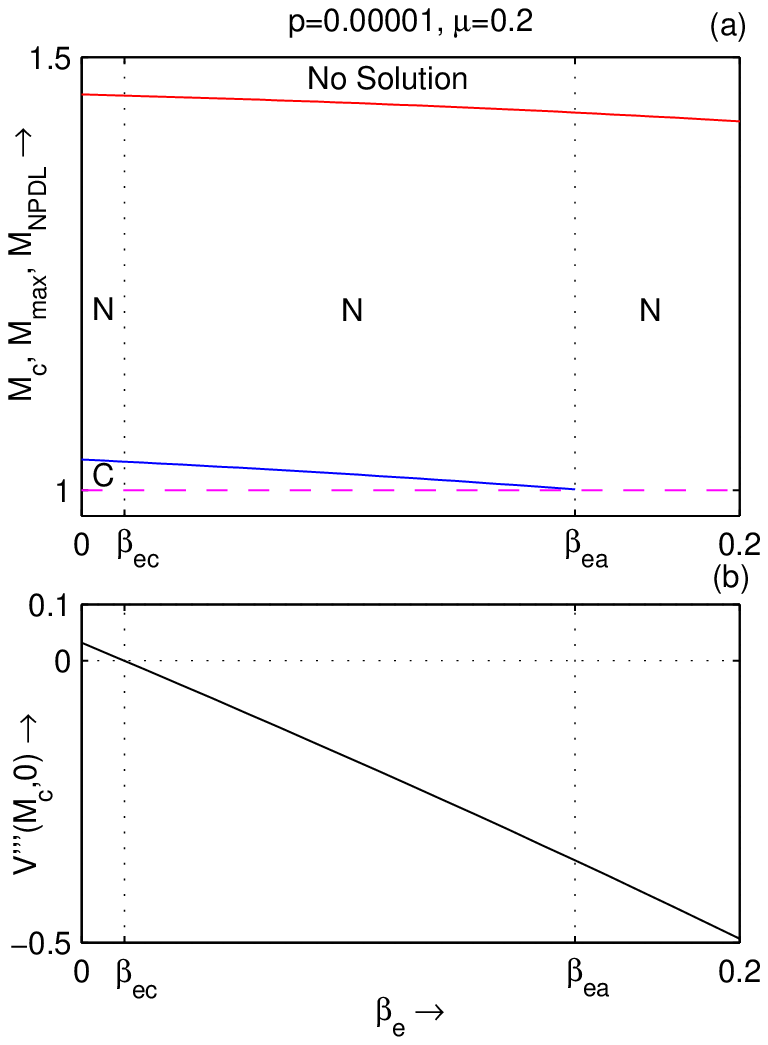}
  \caption{\label{sol_spc_wrt_beta_e_p=0_pt_00001} (a)Existence domain with respect to $\beta_{e}$ for $p=0.00001$, $\mu=0.2$ and $\sigma_{ie}=\sigma_{pe}=0.9$. The red curve, the magenta curve and the blue curve correspond to the curves $M=M_{NPDL}$, $M=M_{c}$ and $M=M_{max}$ respectively. (b) $V'''(M_{c},0)$ is plotted against $\beta_{e}$ for the same values of parameters.}
\end{figure}

\begin{figure}
  \includegraphics{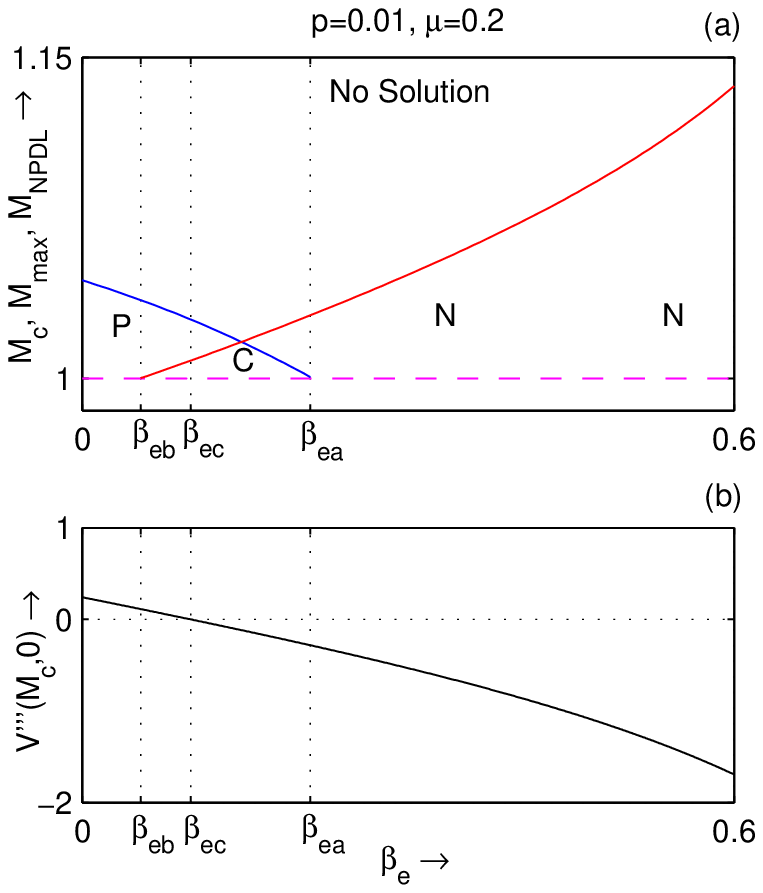}
  \caption{\label{sol_spc_wrt_beta_e_p=0_pt_01} (a)Existence domain with respect to $\beta_{e}$ for $p=0.01$, $\mu=0.2$ and $\sigma_{ie}=\sigma_{pe}=0.9$. The red curve, the magenta curve and the blue curve correspond to the curves $M=M_{NPDL}$, $M=M_{c}$ and $M=M_{max}$ respectively. (b) $V'''(M_{c},0)$ is plotted against $\beta_{e}$ for the same values of parameters.}
\end{figure}

\begin{figure}
  \includegraphics{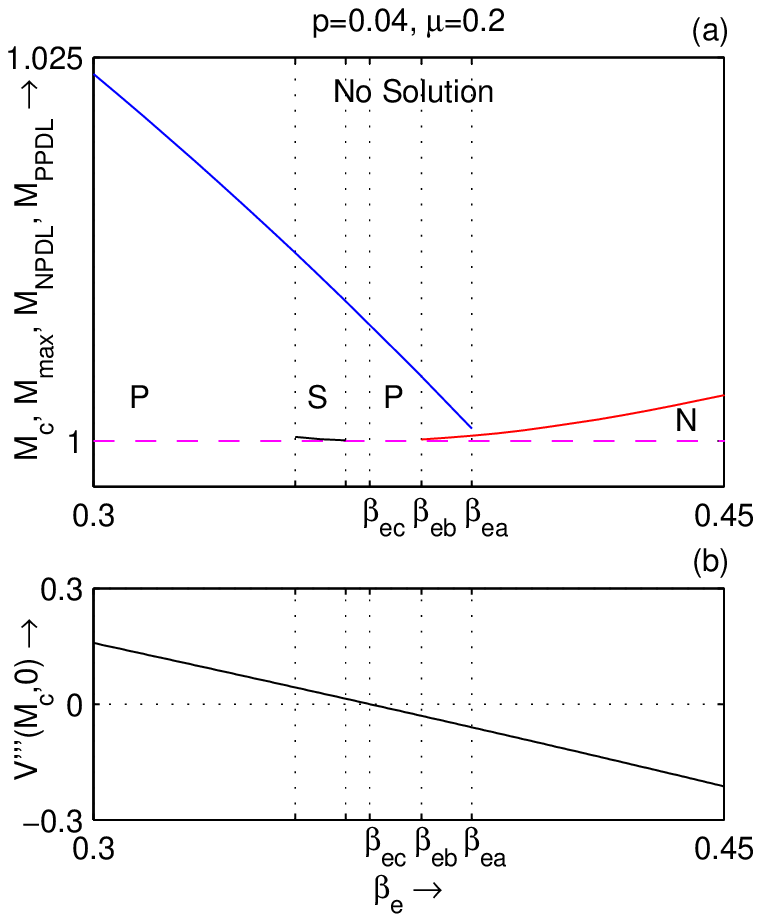}
  \caption{\label{sol_spc_wrt_beta_e_p=0_pt_04} (a)Existence domain with respect to $\beta_{e}$ for $p=0.04$, $\mu=0.2$ and $\sigma_{ie}=\sigma_{pe}=0.9$. The red curve, the magenta curve, the blue curve and the black curve correspond to the curves $M=M_{NPDL}$, $M=M_{c}$, $M=M_{max}$ and $M=M_{NPDL}$ respectively. (b) $V'''(M_{c},0)$ is plotted against $\beta_{e}$ for the same values of parameters.}
\end{figure}

\begin{figure}
\includegraphics{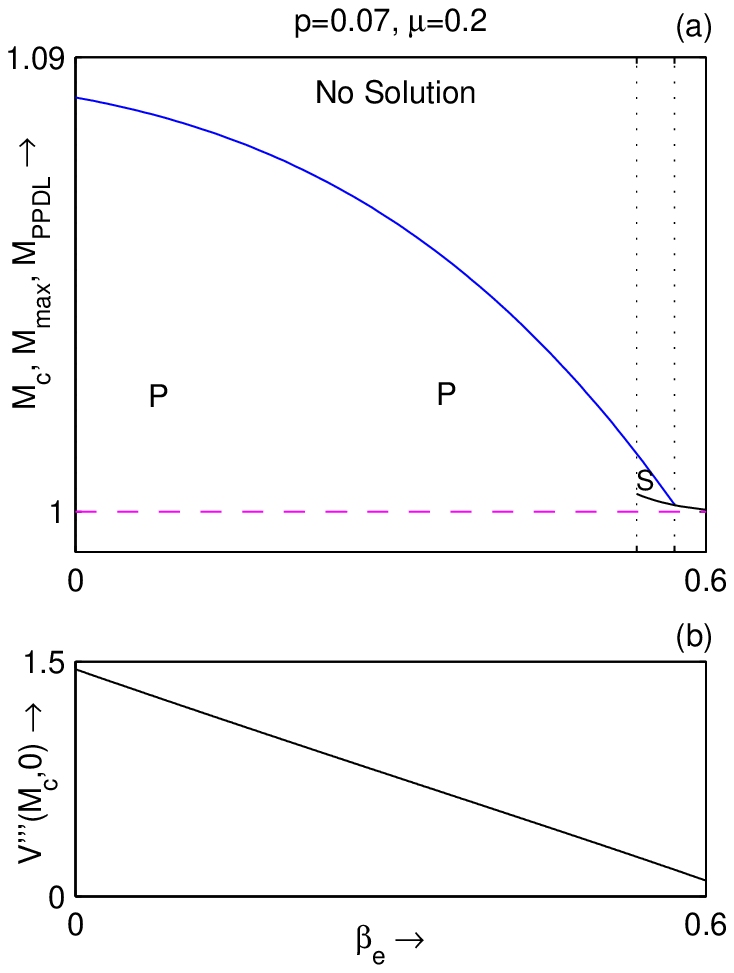}
  \caption{\label{sol_spc_wrt_beta_e_p=0_pt_07} (a)Existence domain with respect to $\beta_{e}$ for $p=0.07$, $\mu=0.2$ and $\sigma_{ie}=\sigma_{pe}=0.9$. The magenta curve, the blue curve and the black curve correspond to the curves $M=M_{c}$, $M=M_{max}$ and $M=M_{NPDL}$ respectively. (b) $V'''(M_{c},0)$ is plotted against $\beta_{e}$ for the same values of parameters.}
\end{figure}

\begin{figure}
\includegraphics{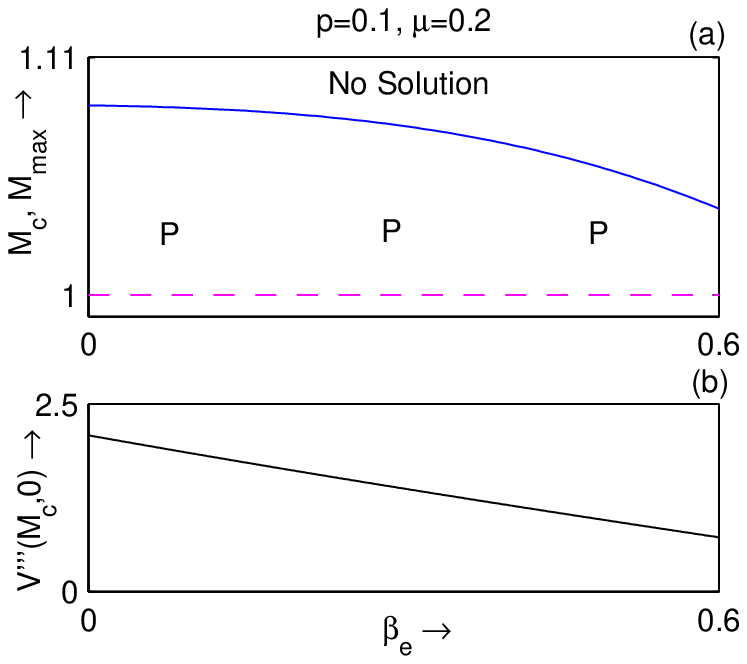}
  \caption{\label{sol_spc_wrt_beta_e_p=0_pt_1} (a)Existence domain with respect to $\beta_{e}$ for $p=0.1$, $\mu=0.2$ and $\sigma_{ie}=\sigma_{pe}=0.9$. The magenta curve and the blue curve correspond to the curves $M=M_{c}$ and $M=M_{max}$ respectively. (b) $V'''(M_{c},0)$ is plotted against $\beta_{e}$ for the same values of parameters.}
\end{figure}

Figures \ref{sol_spc_wrt_beta_e_p=0_pt_00001}(a) - \ref{sol_spc_wrt_beta_e_p=0_pt_1}(a) are qualitatively different existence domains with respect to $\beta_{e}$. These figures show the nature of existence of different solitary structures for $M>M_{c}$ for different values of $p$ whenever $\mu=0.2$ and $\sigma_{ie}=\sigma_{pe}=0.9$.
On the other hand, $V'''(M_{c},0)$ is plotted against $\beta_{e}$ in the lower panel (or marked as (b)) of each figure. 

Although figure \ref{sol_spc_wrt_beta_e_p=0_pt_00001}(a) is the existence domain for $p = 0.00001$, but qualitatively it represents the existence domain for any $p$ lying within the interval $0 <p < 0.0008$ for any physically admissible value of $\beta_{e}$. Similarly, figure \ref{sol_spc_wrt_beta_e_p=0_pt_01}(a), figure \ref{sol_spc_wrt_beta_e_p=0_pt_04}(a) and figure \ref{sol_spc_wrt_beta_e_p=0_pt_07}(a) stand for any $p$ lying within the intervals $0.008 \leq p < 0.034$, $0.034 \leq p <0.065$ and $0.065 \leq p < 0.083$, respectively. Finally, figure \ref{sol_spc_wrt_beta_e_p=0_pt_1}(a) represents the existence domain for $p > 0.083$. 

In the above mentioned figures, P, N, S, and C denote the existence regions of PPSWs, NPSWs, PPSWs after the formation of the PPDL and the region of coexistence of both PPSWs and NPSWs respectively. Here $M_{max}$ is the upper bound of the Mach number $ M $ for the existence of all positive potential solitary structures. Following Das \textit{et al.} \cite{das09,das12},  it is simple to check that $ M_{max} $ is the largest positive root of the equation $ V(\Psi_{M}) = 0 $ subject to the condition $ V(\Psi_{M}) \geq 0 $ for all $ M \leq M_{max} $. In other words, $M$ assumes its upper limit $ M_{max} $ for the existence of all positive potential solitary structures when $\phi$ tends to $\Psi_{M}$, i.e., when ion number density goes to maximum compression. Mach number $M=M_{PPDL}$ ($M=M_{NPDL}$) corresponds to a PPDL (NPDL) solution of the energy integral (\ref{energy_integral}). In our earlier papers \cite{paul17ppr,paul2016,paul17pop}, following Das \textit{et al.} \cite{das09,das12}, we have developed a numerical scheme to find the Mach number $M_{PPDL}$ ($M_{NPDL}$) corresponding to a PPDL (NPDL) solution of the energy integral (\ref{energy_integral}) at some point of the parameter space.      

To investigate the existence and polarity of different solitary structures at $M>M_{c}$, we have defined the following cut off values of $\beta_{e}$:
\begin{description}
  \item[$\beta_{ea}$]  $\beta_{ea}$ is a cut off value of $\beta_{e}$ such that $M_{max}$ (upper bound of the Mach number for the existence of positive potential solitary structures) exists for all $0<\beta_{e} \leq \beta_{ea}$. Consequently, $\beta_{e}=\beta_{ea}$ is the upper bound of $\beta_{e}$ for the existence of PPSWs.
	\item[$\beta_{eb}$] $\beta_{eb}$ is a cut off value of $\beta_{e}$ such that NPDL starts to exist whenever $\beta_{e} \geq \beta_{eb}$ i.e., $\beta_{e}=\beta_{eb}$ is the lower bound of $\beta_{e}$ for the existence of NPDL solution. In other words, for any  $\beta_{e} \geq \beta_{eb}$, there exists a sequence of NPSWs of increasing amplitude which converges to NPDL at $M = M_{NPDL}$.
	\item[$\beta_{ec}$]  $\beta_{ec}$ is the value of $\beta_{e}$ at which $V'''(M_{c},0)=0$.
\end{description}

Now, figuers \ref{sol_spc_wrt_beta_e_p=0_pt_00001} to \ref{sol_spc_wrt_beta_e_p=0_pt_1} are self explanatory. For example, consider figure \ref{sol_spc_wrt_beta_e_p=0_pt_01}. From figure \ref{sol_spc_wrt_beta_e_p=0_pt_01}(a) we see that the system supports only PPSWs in the right neighbourhood of the curve $M=M_{c}$ for $0<\beta_{e}<\beta_{eb}$. Now, NPDLs start to exist for $\beta_{e} \geq \beta_{eb}=0.054$ along the curve $M=M_{NPDL}$ and the coexistence of solitary waves of both polarities has been observed in the interval $\beta_{eb}<\beta_{e} \leq \beta_{ea}$. For $\beta_{e}>\beta_{ea}$, the system supports only NPSWs in the right neighbourhood of the curve $M=M_{c}$. Thus, from figure \ref{sol_spc_wrt_beta_e_p=0_pt_01}(a), we have a clear idea about the existence and polarity of the solitary structures in the right neighbourhood of the curve $M=M_{c}$. On the other hand, figure \ref{sol_spc_wrt_beta_e_p=0_pt_01}(b) shows the variation of $V'''(M_{c},0)$ with respect to $\beta_{e}$. Therefore, using \textbf{Result-1}, \textbf{Result-2} and \textbf{Result-3}, one can draw the following conclusions regarding the existence and the polarity of the solitary structures along the curve $M=M_{c}$.\\
(i) For $0 \leq \beta_{e} < \beta_{eb}$, the system supports only PPSWs in the right neighbourhood of $M=M_{c}$ and in this interval of $\beta_{e}$, we have $V'''(M_{c},0)>0$. So, using \textbf{Result-2} we can conclude that  there does not exist any solitary structure at $M=M_{c}$ for $\beta_{e}$ lying within the interval $0 \leq \beta_{e} < \beta_{eb}$.\\ 
(ii) For $\beta_{eb} < \beta_{e}< \beta_{ec}$, the system supports both PPSWs and NPSWs in the right neighbourhood of $M=M_{c}$ and in this interval of $\beta_{e}$, we have $V'''(M_{c},0)>0$ which indicates the existence of NPSWs at $M=M_{c}$ for $\beta_{eb} < \beta_{e}< \beta_{ec}$ (\textbf{Result-1}).\\
(iii) For $\beta_{ec}  < \beta_{e} \leq \beta_{ea}$, the system again supports both PPSWs and NPSWs in the right neighbourhood of $M=M_{c}$, but in this interval of $\beta_{e}$, we have $V'''(M_{c},0)<0$ which indicates the existence of PPSWs at $M=M_{c}$ in the interval $\beta_{ec}  < \beta_{e} \leq \beta_{ea}$ (\textbf{Result-1}).\\
(iv) For $\beta_{ea}  < \beta_{e} <0.6$, the system supports NPSWs in the right neighbourhood of $M=M_{c}$, but in this interval of $\beta_{e}$, we have $V'''(M_{c},0)<0$. So, we can conclude that  there does not exist any solitary structure at the acoustic speed in the interval $\beta_{ea}  < \beta_{e} <0.6$ (\textbf{Result-2}).\\
(v) It is simple to check that $V'''(M_{c},0)\Big|_{\beta_{e} = \beta_{ec}}=0$ and $V''''(M_{c},0)\Big|_{\beta_{e} = \beta_{ec}}>0$. Therefore, the potential energy of the pseudo particle of unit mass associated with the energy integral (\ref{energy_integral}) attains a minimum value at $\phi=0$ when $\beta_{e} = \beta_{ec}$, $M=M_{c}$ with $p=0.01$, $\mu=0.2$ and $\sigma_{ie}=\sigma_{pe}=0.9$. In this case, $\phi=0$ is the position of stable equilibrium of the particle.   So, it is impossible to make any oscillation of the particle even when the particle is slightly displaced from its stable position of equilibrium and consequently there is no question of the existence of any solitary structure at the acoustic speed $U=C_{D} \Leftrightarrow M=M_{c}$ when $\beta_{e} = \beta_{ec}$.\\
(vi) From figure \ref{sol_spc_wrt_beta_e_p=0_pt_01}(a), we see that the curve $M=M_{NPDL}$ tends to intersect the curve $M=M_{c}$ at the point $\beta_{e} = \beta_{eb}$ in the existence domain with $p=0.01$, $\mu=0.2$ and $\sigma_{ie}=\sigma_{pe}=0.9$, i.e., there always exists a NPDL solution in any right neighborhood of $M_{c}$. Therefore, according to Das \textit{et al.} \cite{das12mc}, there must exist a NPDL solution at $M=M_{c}$ at the point $\beta_{e} = \beta_{eb}$ in the existence domain with $p=0.01$, $\mu=0.2$ and $\sigma_{ie}=\sigma_{pe}=0.9$.\\
(vii) There does not exist any PPDL at the acoustic speed.   

\begin{figure}
  \includegraphics{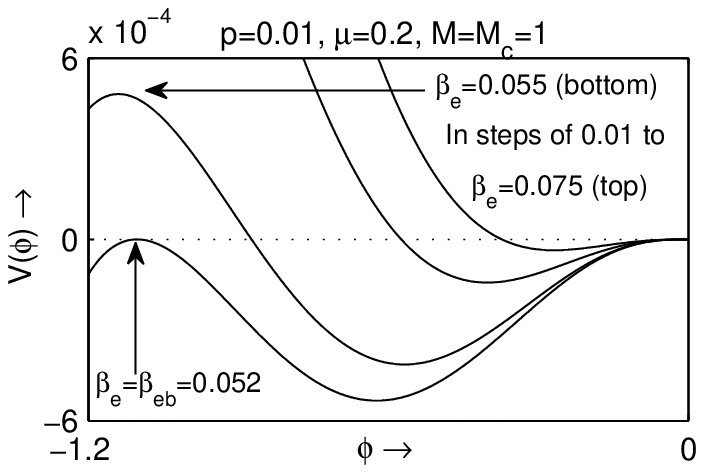}
  \caption{\label{phi_vs_vphi_npsw_p=0_pt_01} $V(\phi)$ is plotted against $\phi$ at $M=M_{c}$ for different values of $\beta_{e}$ and fixed values of other parameters as shown in the figure.}
\end{figure}

\begin{figure}
\includegraphics{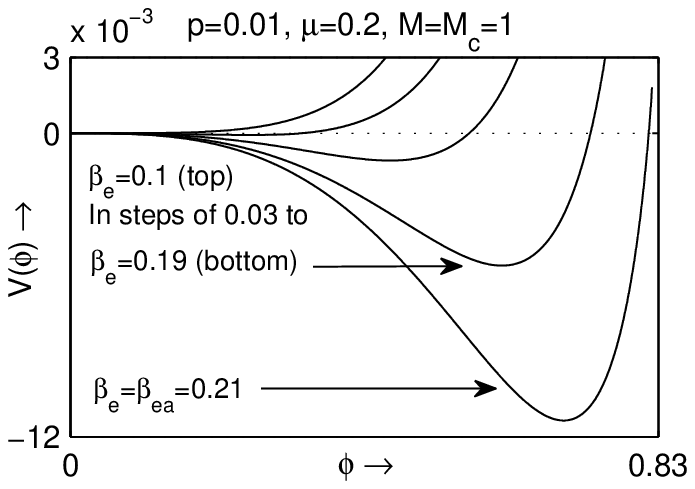}
  \caption{\label{phi_vs_vphi_ppsw_p=0_pt_01} $V(\phi)$ is plotted against $\phi$ at $M=M_{c}$ for different values of $\beta_{e}$ and fixed values of other parameters as shown in the figure.}
\end{figure}

Thus, from figures \ref{sol_spc_wrt_beta_e_p=0_pt_00001} - \ref{sol_spc_wrt_beta_e_p=0_pt_1}, we observe that for $p>0$, there exists a cut off value $p^{(c)}$ of $p$, such that for $0 <p \leq p^{(c)}$,  the system supports NPSWs at the acoustic speed for $0 < \beta_{e}< \beta_{ec}$, whereas for $\beta_{ec}  < \beta_{e} \leq \beta_{ea}$ the system supports PPSWs at the acoustic speed. Again, there exists a cut off value $p^{(k)}$ of $p$, such that for $p^{(c)} < p \leq p^{(k)}$, the system supports NPSWs at the acoustic speed for $\beta_{eb} < \beta_{e}< \beta_{ec}$, whereas for $\beta_{ec}  < \beta_{e} \leq \beta_{ea}$ the system supports PPSWs at the acoustic speed. Here we see that for $M>M_{c}$, the curve $M=M_{NPDL}$ tends to intersect the curve $M=M_{c}$ at $\beta_{e} = \beta_{eb}$ and consequently we have a NPDL solution at the acoustic speed when $\beta_{e} $ assumes the value $\beta_{eb}$. For definiteness, we draw $V(\phi)$ against $\phi$ in figure \ref{phi_vs_vphi_npsw_p=0_pt_01} at the acoustic speed for different values of $\beta_{e} $ lying in the interval $\beta_{eb} \leq \beta_{e}< \beta_{ec}$. From this figure we see that the amplitude of the NPSWs at the acoustic speed increases with decreasing $\beta_{e} $ and this sequence of NPSWs ends with a NPDL at $\beta_{e} = \beta_{eb}$. In figure \ref{phi_vs_vphi_ppsw_p=0_pt_01}, $V(\phi)$ is plotted against $\phi$ for different values of $\beta_{e} $ lying in the interval $\beta_{ec}  < \beta_{e} \leq \beta_{ea}$. From this figure we see that the amplitude of the PPSWs at $M=M_{c}$ increases with increasing $\beta_{e} $ lying within the interval $\beta_{ec}  < \beta_{e} \leq \beta_{ea}$, whereas at the point $\beta_{e} = \beta_{ec}$ both NPSWs and PPSWs collapse. It is simple to check that potential energy of the system assumes a minimum value when $\beta_{e} = \beta_{ec}$, i.e., at $\beta_{e} = \beta_{ec}$, $\phi=0$ is a position of stable equilibrium. In fact, at this point $V(M_{c},0)=V'(M_{c},0)=V''(M_{c},0)=V'''(M_{c},0)=0$, whereas  $V''''(M_{c},0)>0$. Consequently, there is no question of the existence of any solitary structure at $\beta_{e} = \beta_{ec}$. For further increment in $p$ from $p=p^{(k)}$, there exists a cut off value $p^{(m)}$ of $p$, such that for $p^{(k)} < p \leq p^{(m)}$, the system supports PPSWs at the acoustic speed for $\beta_{ec} < \beta_{e}< \beta_{eb}$, whereas the system does not support any negative potential solitary structure at the acoustic speed for any admissible value of $\beta_{e} $.
 Finally, for $p>p^{(m)}$, the system does not support any solitary structure at $M=M_{c}$. Again, from figure \ref{sol_spc_wrt_beta_e_p=0_pt_04} and figure \ref{sol_spc_wrt_beta_e_p=0_pt_07}, we see that the system supports PPDLs in a right neighbourhood of $M=M_{c}$, but the system does not support PPDLs at $M=M_{c}$.  Consequently, the present plasma system does not support any positive potential supersoliton at the acoustic speed. Again, since there does not exist any soliton after the formation of NPDL at the acoustic speed, there does not exist any negative potential supersoliton at the acoustic speed for $p>0$. But for $p=0$, i.e., if there is no positrons in the system, Das \textit{et al.} \cite{das2011existence} found the existence of NPDLs and most importantly, the existence of negative potential supersolitons at the acoustic speed. Now, in the present work, if we consider $p=0$, i.e., the positron concentration in the system is zero, then the present plasma system reduces to the exactly same system of Das \textit{et al.} \cite{das2011existence} 
 
\begin{figure}
  \includegraphics{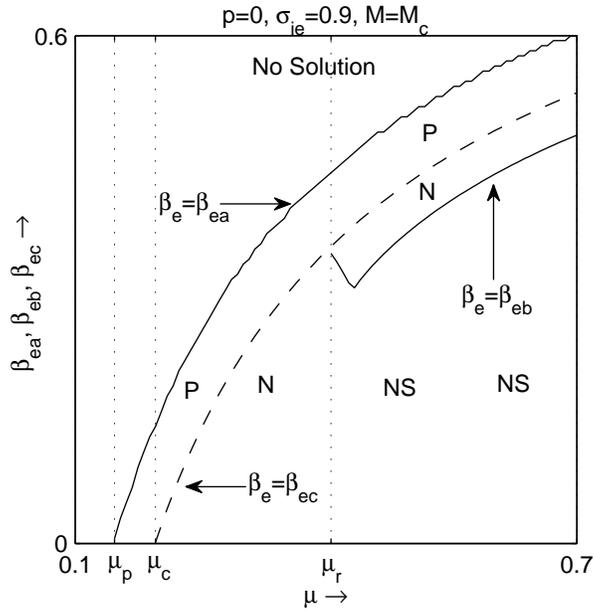}
  \caption{\label{sol_spc_wrt_mu_@M=Mc_p=0} Existence domain with respect to $\mu$ at $M=M_{c}$ for different values of parameters as shown in the figure.}
\end{figure}

Figure \ref{sol_spc_wrt_mu_@M=Mc_p=0} is the existence domain with respect to $\mu$ at $p=0$ when $\sigma_{ie}=0.9$. This existence domain is qualitatively same as the existence domain as given in figure 10 in the paper of Das \textit{et al.} \cite{das2011existence} In figure \ref{sol_spc_wrt_mu_@M=Mc_p=0}, P stands for the existence region of PPSWs at $M=M_{c}$, N stands for the existence region of NPSWs at $M=M_{c}$, along the curve $\beta_{e}=\beta_{eb}$ we have NPDLs at $M=M_{c}$, NS represents the existence region of NPSWs after the formation of NPDL at $M=M_{c}$ and $V'''(M_{c},0)=0$ along the curve $\beta_{e}=\beta_{ec}$. This figure shows the existence domain with respect to $\mu$ at the acoustic speed $U=C_{D} \Leftrightarrow M=M_{c}$ for $p=0$ and $\sigma_{ie}=0.9$. To describe figure \ref{sol_spc_wrt_mu_@M=Mc_p=0}, we have defined the following cut off values of $\mu$:
\begin{description}
	\item[$\mu_{p}$] $\mu_{p}$ is a cut off value of $\mu$ such that $M_{max}$ does not exist for any admissible value of $\beta_{e}$ if $\mu$ lies within the interval $0 < \mu < \mu_{p}$, i.e., if $\mu \geq \mu_{p}$, there exists a value $\beta_{e}^{*}$ of $\beta_{e}$ such that $M_{max}$ exists at $\beta_{e}=\beta_{e}^{*}$; moreover, if $\beta_{e}^{*}>0$, then $M_{max}$ exists for all $\beta_{e}$ lies within the interval $0 \leq \beta_{e} < \beta_{e}^{*}$.
	\item[$\mu_{c}$] $\mu_{c}$ is a cut off value of $\mu$ such that $V''(M_{c},0)=0$ and $V'''(M_{c},0)=0$ at $\mu=\mu_{c}$ and $\beta_{e}=\beta_{ec}$ for fixed values of $\sigma_{ie}$.
	\item[$\mu_{r}$] $\mu_{r}$ is another cut off value of $\mu$ such that for all $\mu_{r} \leq \mu_{T}$, the curve $M = M_{NPDL}$ tends to intersect the curve $M = M_{c}$ at the point $\beta_{e}=\beta_{eb}$.
	\item[$\mu_{T}$] $\mu_{T}$ is a physically admissible upper bound of $\mu$.
\end{description}
 
The Figure \ref{sol_spc_wrt_mu_@M=Mc_p=0} clearly shows that there exist two types of NPSWs at $M=M_{c}$ if $\mu$ lies within the interval $\mu_{r}< \mu < \mu_{T}$. The first type is bounded by the curves $\beta_{e}=\beta_{ec}$ and $\beta_{e}=\beta_{eb}$ and the amplitude of these NPSWs is restricted by the amplitude of NPDL at the acoustic speed. The second type of NPSWs exist beyond the curve $\beta_{e}=\beta_{eb}$, i.e., after the formation of NPDL at $M=M_{c}$. The amplitude of the NPSWs after the formation of double layer at $M=M_{c}$ increases with decreasing $\beta_{e}$ and finally, attains its maximum value when $\beta_{e}=0$ at $M=M_{c}$. Again, there exists a jump type discontinuity between the amplitude of the NPSWs at the acoustic speed just before and after the formation of NPDL at $M=M_{c}$ (see figure \ref{profile_supersoliton_p=0}). Since, the existence of solitons after the formation of double layer confirms the existence of at least one sequence of supersolitons \cite{paul17pop}, therefore, we can conclude that whenever there exists no positron in the system, it supports negative potential supersolitons at the acoustic speed. Consequently, there must be a smooth transition of solitary structures at $M=M_{c}$, viz., soliton $\to$ double layer $\to$ supersoliton $\to$ soliton. For the first time, this transition process has been elaborately discussed in the next section with the help of phase portraits of the dynamical system corresponding to the DIA solitary structures at the acoustic speed.


\section{\label{sec:Phase_Portraits} Phase Portraits of different solitary structures at the acoustic speed}

Before going to investigate the mechanism of transition of the solitary structures at the acoustic speed, we have to describe the phase portraits of the dynamical system corresponding to the different solitary structures at the acoustic speed. It is also necessary to make a clear difference between the solitary structures at the acoustic speed, i.e., at $M=M_{c}$ and the solitary structures at the supersonic speed, i.e., for $M>M_{c}$.

Differentiating the energy integral (\ref{energy_integral}) with respect to $\phi$, we get the following differential equation:
\begin{eqnarray}\label{energy_integral_differentiation}
\frac{d^{2}\phi}{d\xi^{2}}+V'(\phi)=0.
\end{eqnarray}
This equation is equivalent to the following system of differential equations
\begin{eqnarray}\label{phase_portraits}
\frac{d\phi_{1}}{d\xi}=\phi_{2}~,~\frac{d\phi_{2}}{d\xi}=-V'(\phi_{1})~,
\end{eqnarray}
where $\phi_{1}=\phi$. In the present paper, we have considered the solitary structures at $M=M_{c}$ with the help of qualitatively different existence domains. Now, we explain their shapes with the help of phase portraits of the system of coupled equations (\ref{phase_portraits}) in the $\phi_{1}-\phi_{2}$ plane.

The fixed point of the dynamical system (\ref{phase_portraits}) is ($\phi_{1}^{*}$, $\phi_{2}^{*}$), where $\phi_{2}^{*}=0$ and $\phi_{1}^{*}$ is given by the equation
	\begin{eqnarray}\label{phi_1_star}
		V'(\phi_{1}^{*})=0
	\end{eqnarray}  
	  This equation gives the value(s) of $\phi_{1}^{*}$ as a function of the physical parameters of the system at the Mach number $M=M_{c}=1$, i.e., $\phi_{1}^{*}$ is a function of $p$, $\mu$ $\beta_{e}$, $\sigma_{ie}$ and $\sigma_{pe}$. So, we can write
	\begin{eqnarray}\label{phi_1_star_function}
		\phi_{1}^{*}=\phi_{1}^{*}(p,\mu,\beta_{e},\sigma_{ie},\sigma_{pe}).
	\end{eqnarray}
In the present work, we take $\sigma_{pe}=0.9$, i.e., the average thermal temperatures of positrons is nearly same as that of electrons and we have also considered $\sigma_{ie}=0.9$ (the usual dusty plasma approximation $T_{i}\approx T_{e}$). Therefore, for fixed values of $p$ and $\mu$, the equation (\ref{phi_1_star_function}) reduces to
\begin{eqnarray}\label{phi_1_star_function_1}
		\phi_{1}^{*}=\phi_{1}^{*}(\beta_{e}).
	\end{eqnarray}
To know the value of  $\beta_{e}$, we have already drawn the existence domains with respect to $\beta_{e}$ (see figure \ref{sol_spc_wrt_beta_e_p=0_pt_00001}(a), \ref{sol_spc_wrt_beta_e_p=0_pt_01}(a), \ref{sol_spc_wrt_beta_e_p=0_pt_04}(a),
\ref{sol_spc_wrt_beta_e_p=0_pt_07}(a), and \ref{sol_spc_wrt_beta_e_p=0_pt_1}(a)) and from these existence domains, we can easily decide the value of $\beta_{e}$ for the existence of the desired solitary structure at the acoustic speed.

\begin{figure}
\begin{center}
\includegraphics{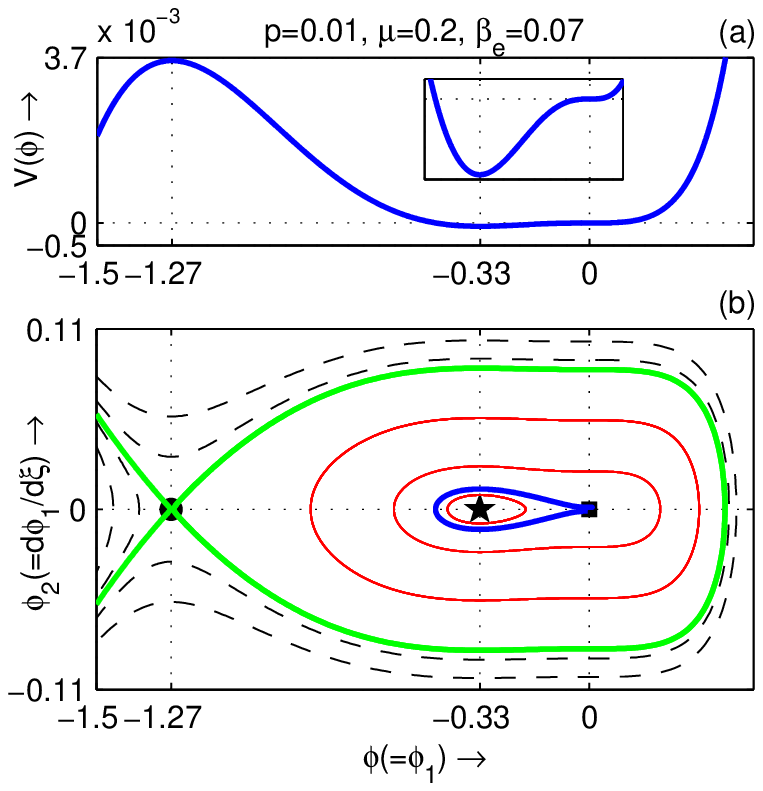}
  \caption{\label{pp_npsw_p=0_pt_01} $V(\phi)$ (on top) and the phase portrait of the system (\ref{phase_portraits}) (on bottom) have been drawn on the same $\phi(=\phi_{1})$-axis at $M=M_{c}$ when $p=0.01$, $\mu=0.2$, $\beta_{e}=0.07$ and $\sigma_{ie}=\sigma_{pe}=0.9$.}
\end{center}
\end{figure}

\begin{figure}
\begin{center}
\includegraphics{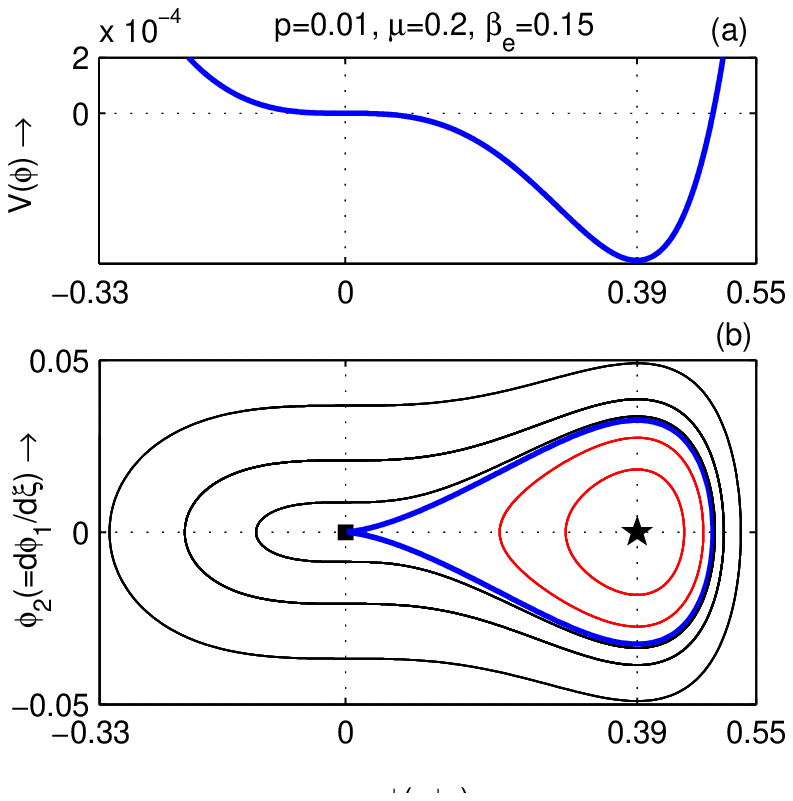}
  \caption{\label{pp_ppsw_p=0_pt_01} $V(\phi)$ (on top) and the phase portrait of the system (\ref{phase_portraits}) (on bottom) have been drawn on the same $\phi(=\phi_{1})$-axis at $M=M_{c}$ when $p=0.01$, $\mu=0.2$, $\beta_{e}=0.15$ and $\sigma_{ie}=\sigma_{pe}=0.9$.}
\end{center}
\end{figure}

\begin{figure}
\begin{center}
\includegraphics{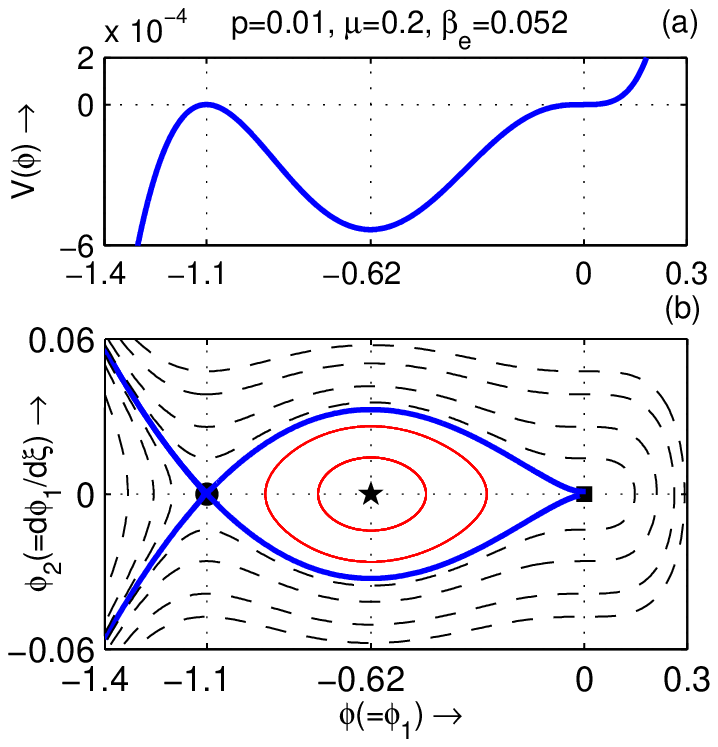}
  \caption{\label{final_pp_NPDL_p=0_pt_01} $V(\phi)$ (on top) and the phase portrait of the system (\ref{phase_portraits}) (on bottom) have been drawn on the same $\phi(=\phi_{1})$-axis at $M=M_{c}$ when $p=0.01$, $\mu=0.2$, $\beta_{e}=0.07$ and $\sigma_{ie}=\sigma_{pe}=0.9$.}
\end{center}
\end{figure}

To describe the phase portraits of the solitary structures at $M=M_{c}$, we consider figures \ref{pp_npsw_p=0_pt_01} - \ref{final_pp_NPDL_p=0_pt_01}. Here we have used the existence domain as shown in figure \ref{sol_spc_wrt_beta_e_p=0_pt_01} to determine the value of $\beta_{e}$ for the existence of desired solitary structure at the acoustic speed. In figures \ref{pp_npsw_p=0_pt_01}(a) - \ref{final_pp_NPDL_p=0_pt_01}(a), $V(\phi)$ is plotted against $\phi$. The lower panel (or marked as (b)) of each figure shows the phase portrait of the system (\ref{phase_portraits}). In these figures, we have used the values of the parameters as indicated in the figures with $p=0.01$, $\mu=0.2$ and $\sigma_{pe}=\sigma_{ie}=0.9$. The curve $V(\phi)$ and the  phase portrait have been drawn on the same horizontal axis $\phi(=\phi_{1})$. The small solid square corresponds to the point of inflexion at the origin, the small solid circle corresponds to a saddle point and the small solid star indicates an equilibrium point other than saddle point or the point of inflexion of the system (\ref{phase_portraits}). It is simple to check that each maximum (minimum) point of $V(\phi)$ corresponds to a saddle point (an equilibrium point other than a saddle point) of the system (\ref{phase_portraits}). Again, small solid square corresponds to the point of inflexion of the system (\ref{phase_portraits}). The concept of the point of inflexion in the study of solitons at the acoustic speed is not new one. Das \textit{et al.} \cite{das12mc} have already mentioned that the origin is the point of inflexion of the system (\ref{phase_portraits}) for solitary structures at the acoustic speed. In fact, if $V(0)=V'(0)=0$, $V''(M_{c},0)=0$ and $V'''(M_{c},0)\neq 0$, the point $\phi=0$ is the point of inflexion which seperates the convex part of the curve $V(M_{c},\phi)$ from its concave part. According to Theorms 3 and 4 of Das \textit{et al.} \cite{das12mc}, the origin $(0,0)$ is always a point of inflexion of the system (\ref{phase_portraits}) for solitary structures at the acoustic speed. But in case of supersonic solitary structures ($M>M_{c}$), the origin $(0,0)$ is not the point of inflexion of the system (\ref{phase_portraits}). In this case, i.e., for the supersonic case, the origin $(0,0)$ is always a saddle point of the system (\ref{phase_portraits}). This gives a difference between the solitary structures for $M>M_{c}$ and the solitary structures at $M=M_{c}$. 

Now, there is a one-one correspondence between the separatrix of the phase portrait as shown with a heavy blue line in the lower panel with the curve $V(\phi)$ against $\phi$ of the upper panel. In fact, this one-one correspondence between the separatrix of the phase portrait and the curve $V(\phi)$ against $\phi$ has been elaborately discussed by Paul \textit{et al.} \cite{paul17pop} for supersonic solitary structures. In this section, we want to discuss the phase portraits of the solitary structures at the acoustic speed and the transition process: solitons $\to$ double layers $\to$ supersolitons $\to$ solitons after the formation of double layer  at the acoustic speed when there is no positrons in the system. For the sonic case, i.e., at the acoustic speed, the separatrix corresponding to a solitary structure starts from the point of inflexion (0,0) and ends at the point of inflexion (0,0). This shows that if a separatrix is formed in the positive $\phi$ - axis then it is impossible to form another separatrix in the negative direction of $\phi$ - axis, and consequently coexistence of solitary structures of both polarities is not possible at the acoustic speed. This is also not a new result because Das \textit{et al.} \cite{das12mc} have already proved the following theorem: \textit{\textbf{Theorem 5:} If $V(0)=V'(0)=0$, $V''(M_{c},0)=0$ and $V'''(M_{c},0)\neq 0$, it is not possible to have coexistence of both positive and negative potential solitary structures at $M=M_{c}$.} Therefore, phase portrait analysis confirms the \textbf{Result - 3} or \textbf{Theorem -5}. The separatrix corresponding to a solitary structure is shown with a heavy blue line, whereas other separatrices (if exist) are shown by green lines. The closed curve about an equilibrium point (other than a saddle point or the point of inflexion) contained in at least one separatrix indicates the possibility of the periodic wave solution about that fixed point.

Figure \ref{pp_npsw_p=0_pt_01}(a) shows the existence of a NPSW at $M=M_{c}$ and figure \ref{pp_npsw_p=0_pt_01}(b) describes the corresponding phase portrait. Here we see that the system has a point of inflexion at the origin, an equilibrium point at $(-0.33,0)$ and a saddle at $(-1.27,0)$. Again, from figure \ref{pp_npsw_p=0_pt_01}(b), we see that there are two separatrices: (i) the separatrix (as shown by a heavy blue line) that starts and ends at the origin enclosing the non-saddle fixed point and this separatrix corresponds to the negative potential soliton at $M=M_{c}$ and (ii) the separatrix (as shown by heavy green line) which appears to pass through the saddle point $(-1.27,0)$ and this separatrix contains the separatrix (as shown by a heavy blue line) that starts and ends at the origin. There exist infinitely many closed curves between these two separatrices and each of these closed curves corresponds to a super-nonlinear periodic wave. Thus, figure \ref{pp_npsw_p=0_pt_01}(b) confirms the existence of super-nonlinear periodic waves at the acoustic speed.

Figure \ref{pp_ppsw_p=0_pt_01}(a) shows the existence of a PPSW at $M=M_{c}$ and figure \ref{pp_ppsw_p=0_pt_01}(b) describes the corresponding phase portrait. Here we see that the system has a point of inflexion at the origin, an equilibrium point at $(0.39,0)$ which is not a saddle point. From figure \ref{pp_ppsw_p=0_pt_01}(b), we see that there exists only one separatrix (as shown by a heavy blue line) that starts and ends at the origin enclosing the non-saddle fixed point $(0.39,0)$ and consequently this separatrix corresponds to the positive potential soliton at $M=M_{c}$.

Figure \ref{final_pp_NPDL_p=0_pt_01}(b) shows the phase portrait of a NPDL at the acoustic speed and this figure shows that the separatrix corresponding to the double layer solution at the acoustic speed appears to start and end at the point of inflexion $(0,0)$ and again it appears to pass through the saddle point at $(-1.1,0)$ enclosing the non-saddle fixed point $(-0.62,0)$. In figure \ref{final_pp_NPDL_p=0_pt_01}(a), $V(\phi)$ is plotted against $\phi$ at the acoustic speed for the given values of the parameters as indicated in the figure. Figure \ref{final_pp_NPDL_p=0_pt_01}(a) and figure \ref{final_pp_NPDL_p=0_pt_01}(b) together give a one-one correspondence between the separatrix of the phase portrait as shown with a heavy blue line in the lower panel with the curve $V(\phi)$ against $\phi$ of the upper panel. This mechanism holds good for formation of PPSWs and also for the formation of NPSWs at the acoustic speed.

\begin{figure}
\includegraphics{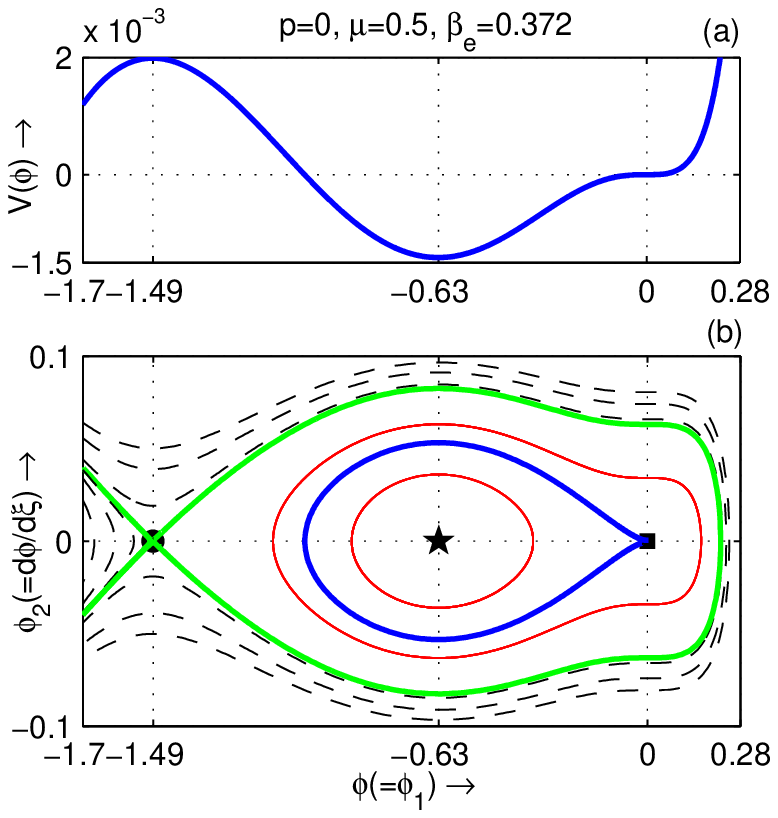}
  \caption{\label{pp_npsw_p=0} $V(\phi)$ (on top) and the phase portrait of the system (\ref{phase_portraits}) (on bottom) have been drawn on the same $\phi(=\phi_{1})$-axis at $M=M_{c}$ when $p=0$, $\mu=0.5$, $\beta_{e}=0.372$ and $\sigma_{ie}=0.9$.}
\end{figure}

\begin{figure}
\begin{center}
\includegraphics{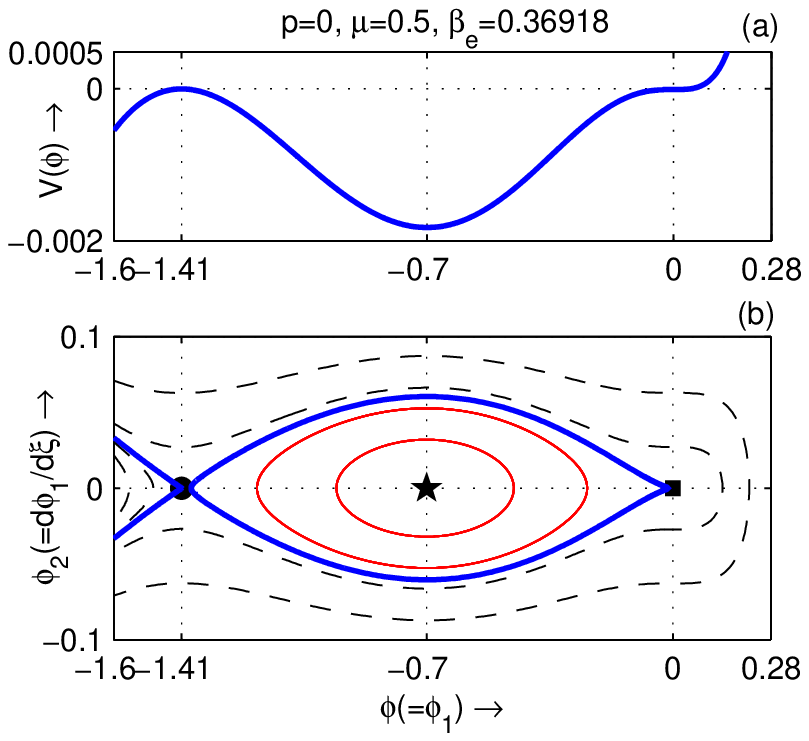}
  \caption{\label{pp_npdl_p=0} $V(\phi)$ (on top) and the phase portrait of the system (\ref{phase_portraits}) (on bottom) have been drawn on the same $\phi(=\phi_{1})$-axis at $M=M_{c}$ when $p=0$, $\mu=0.5$, $\beta_{e}=0.36918$ and $\sigma_{ie}=0.9$.}
\end{center}
\end{figure}

\begin{figure}
\begin{center}
\includegraphics{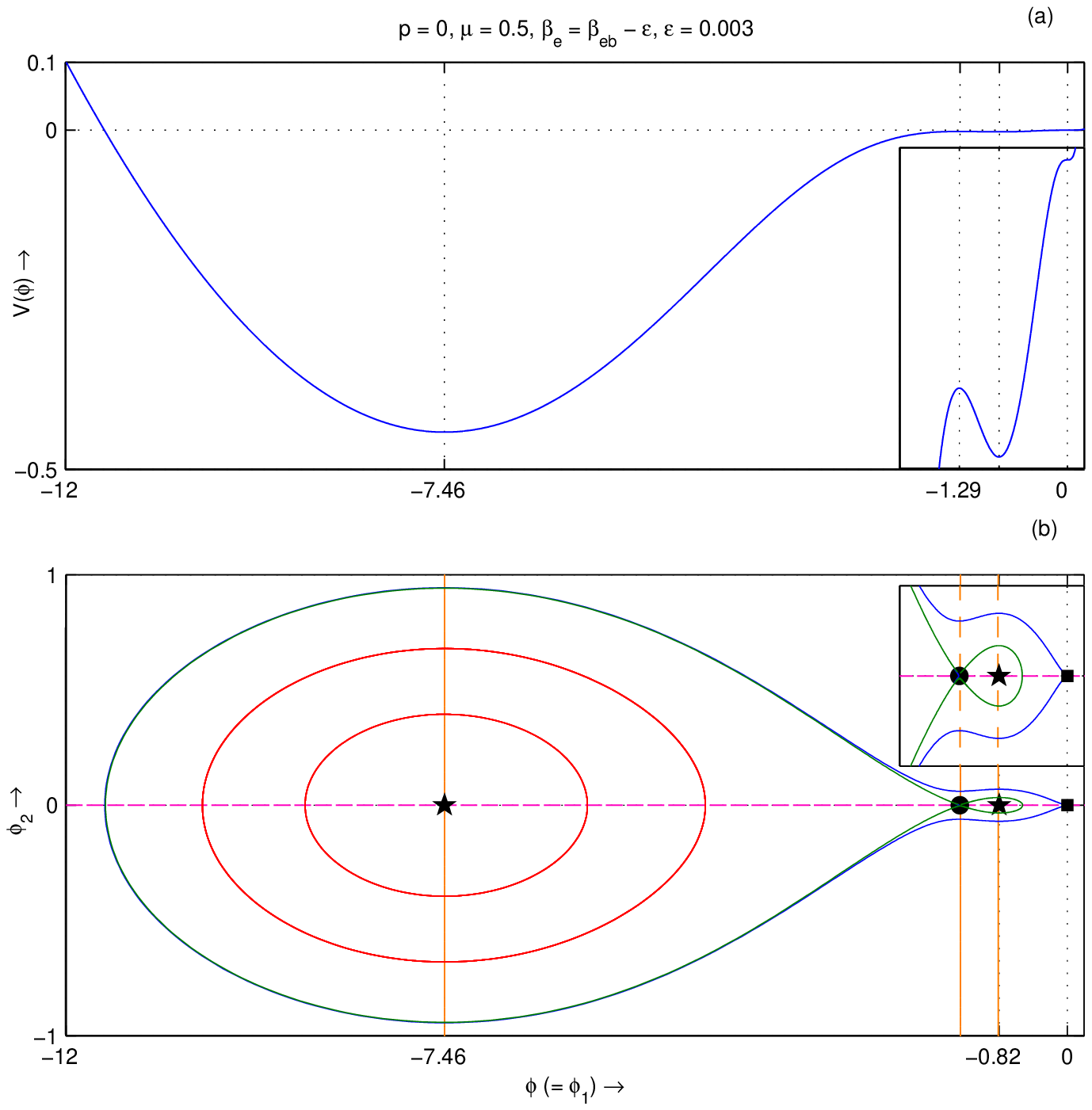}
  \caption{\label{pp_npsupersoliton_p=0} $V(\phi)$ (on top) and the phase portrait of the system (\ref{phase_portraits}) (on bottom) have been drawn on the same $\phi(=\phi_{1})$-axis at $M=M_{c}$ when $p=0$, $\mu=0.5$, $\beta_{e}=0.36908$ and $\sigma_{ie}=0.9$.}
\end{center}
\end{figure}

\begin{figure}
\begin{center}
\includegraphics{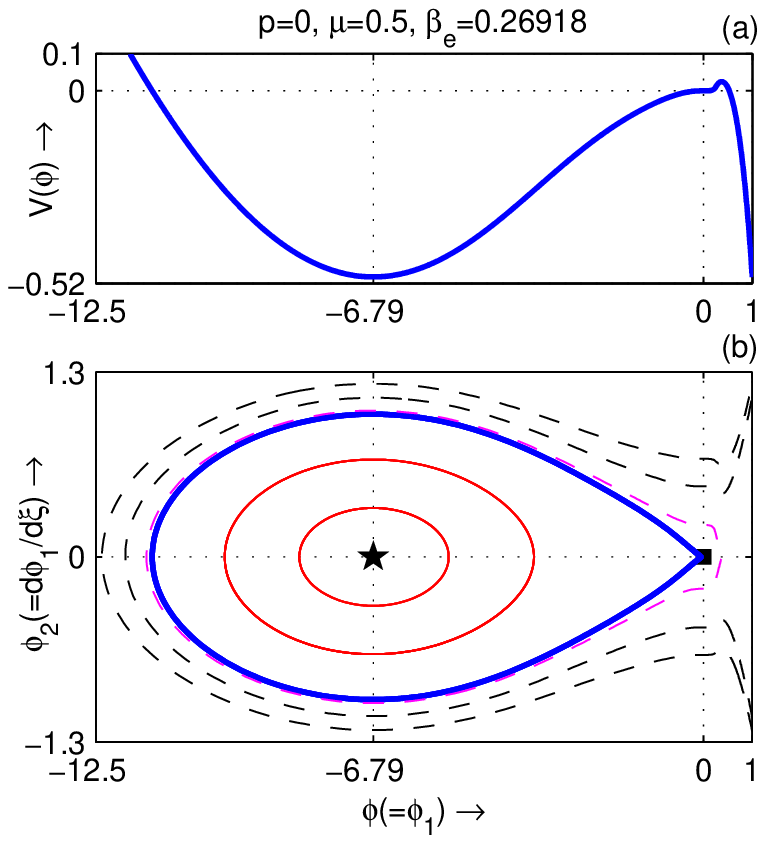}
  \caption{\label{pp_npsw_after_npdl_p=0} $V(\phi)$ (on top) and the phase portrait of the system (\ref{phase_portraits}) (on bottom) have been drawn on the same $\phi(=\phi_{1})$-axis at $M=M_{c}$ when $p=0$, $\mu=0.5$, $\beta_{e}=0.26918$ and $\sigma_{ie}=0.9$.}
\end{center}
\end{figure}

\begin{figure}
\begin{center}
\includegraphics{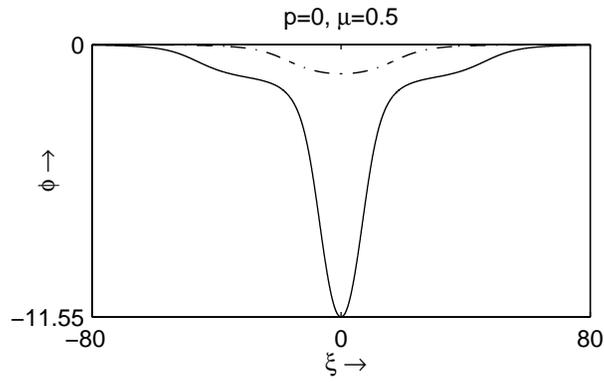}
  \caption{\label{profile_supersoliton_p=0} $\phi$ is plotted against $\xi$ at $M=M_{c}$ for $\beta_{e}=0.26918$ (solid curve) and $\beta_{e}=0.372$ (dash-dot curve) when $p=0$, $\mu=0.5$ and $\sigma_{ie}=0.9$.}
\end{center}
\end{figure}

Now, we are in a position to discuss the transition process of the solitary structures, viz., solitons $\to$ double layers $\to$ supersolitons $\to$ solitons after the formation of double layer  at the acoustic speed when there is no positrons in the system. In this case, i.e., for $p=0$, we consider the existence domain as given in figure \ref{sol_spc_wrt_mu_@M=Mc_p=0} to find the values of $\mu$ and $\beta_{e}$ for the existence of the desired solitary structure at the acoustic speed. One - one correspondence between the separatrix of the phase portrait with the curve $V(\phi)$ against $\phi$ and the transition process of different solitary structures at the acoustic speed have been shown through the figures \ref{pp_npsw_p=0} - \ref{equilibrium_points_p=0}.

Figure \ref{pp_npsw_p=0}(a) shows the existence of a NPSW at $M=M_{c}$ before the formation of NPDL and figure \ref{pp_npsw_p=0}(b) describes the phase portrait of the dynamical system (\ref{phase_portraits}) at $M=M_{c}$ for the values of the parameters as mentioned in the figure. Here we see that the system has a point of inflexion at the origin, an equilibrium point at $(-0.63,0)$ and a saddle at $(-1.49,0)$. Again, from figure \ref{pp_npsw_p=0}(b), we see that there are two separatrices. (i) The separatrix (as shown by a heavy blue line) that appears to start and end at the origin enclosing the non-saddle fixed point corresponds to the negative potential soliton at $M=M_{c}$. (ii) This blue separatrix is contained in another separatrix (as shown by a heavy green line) that appears to pass through the saddle point $(-1.49,0)$. There exist infinitely many closed curves between these two separatrices and each of these closed curves corresponds to a super-nonlinear periodic wave. Thus, figure \ref{pp_npsw_p=0}(b) confirms the existence of super-nonlinear periodic waves at the acoustic speed.

Figure \ref{pp_npdl_p=0}(b) shows the phase portrait of NPDL at the acoustic speed when $\beta_{e}=\beta_{eb}=0.36918$  and this figure shows that the separatrix corresponding to the double layer solution at the acoustic speed appears to start and end at the point of inflexion $(0,0)$ and again it appears to pass through the saddle point at $(-1.41,0)$ enclosing the non-saddle fixed point $(-0.7,0)$. Now we slightly decrease the value of $\beta_{e}$ from $\beta_{eb}$ and draw figure \ref{pp_npsupersoliton_p=0} for $\beta_{e}=\beta_{eb}-0.003$. In figure \ref{pp_npsupersoliton_p=0}(a), $V(\phi)$ is plotted against $\phi$, where the region between $-2$ and $0$ is shown in larger scale in the inset, whereas figure \ref{pp_npsupersoliton_p=0}(b) describes the phase portrait of the dynamical system (\ref{phase_portraits}) at $M=M_{c}$ for the values of the parameters as mentioned in the figure, where the region between $-2$ and $0$ is shown in larger scale in the inset. The figure \ref{pp_npsupersoliton_p=0}(a) shows that $V(\phi)$ has two consecutive minima at $\phi = -0.81504 \approx -0.82$ and at $\phi = -7.4622 \approx -7.46$. Consequently, the phase portrait of the system has two non-saddle fixed points as shown in the lower panel of figure \ref{pp_npsupersoliton_p=0}. The separatrix crresponding to the solitary structure that appears to start and end at the point of inflexion $(0,0)$ encloses one non-zero saddle point $(-1.2875,0)\approx (-1.29,0)$, and two non-saddle fixed points $(-0.81504,0)\approx (-0.82,0)$ and $(-7.4622,0)\approx(-7.46,0)$. From the region between $-2$ and $0$ as shown in larger scale in the inset of figure \ref{pp_npsupersoliton_p=0}(b), one can also check that this separatrix also envelopes one inner separatrix (shown by a green line) that appears to pass through the saddle point $(-1.2875,0)\approx (-1.29,0)$. Therefore, according to the definition of supersoliton, this separatrix is associated with a new type of solitary wave at the acoustic speed - a supersoliton at the acoustic speed. Thus, figure \ref{pp_npsupersoliton_p=0} confirms the existence of negative potential supersolitons at the acoustic speed. Now, we further reduce the value of $\beta_{e}$ and draw figure \ref{pp_npsw_after_npdl_p=0} for $\beta_{e}=\beta_{eb}-0.1=0.26908$. From figure \ref{pp_npsw_after_npdl_p=0}(b), we see that the phase portrait is qualitatively same as the phase portrait of NPSW at the acoustic speed (as shown in \ref{pp_npsw_p=0}(b)). Although, there exists a jump type discontinuity between the amplitudes of the solitons before and after the fomation of NPDL at the acoustic speed (see figure \ref{profile_supersoliton_p=0}). Therefore, for decreasing values of $\beta_{e}$ with $\beta_{e}<\beta_{eb}$, the negative potential supersolitons ultimately reduce to NPSWs after the formation of NPDL. In other words, there must be a critical value $\beta_{e}^{(cr)}$ of $\beta_{e}$, such that the system supports negative potential supersoliton at the acoustic speed when $\beta_{e}$ lies within the interval $\beta_{eb} < \beta_{e} < \beta_{e}^{(cr)}$ and for $\beta_{e} > \beta_{e}^{(cr)}$, the system supports NPSW after the formation of double layer at the acoustic speed. Thus, we see that there exists a transition between the solitary structures at the acoustic speed, viz., soliton $\to$ double layer $\to$ supersoliton $\to$ soliton after the formation of double layer . Such transition process have also been observed by Paul \textit{et al.} \cite{paul17pop} for the solitary structures in case of supersonic waves, i.e., for $M>M_{c}$. To understand the mechanism of this transition process of solitary structures at the acoustic speed, we plot the origin (i.e., the point of inflexion), the saddle and other equilibrium points of the system (\ref{phase_portraits}) on the $\phi(=\phi_{1})$-axis for decreasing values of $\beta_{e}$ starting from $\beta_{e}=\beta_{eb}-0.000001$ in figure \ref{equilibrium_points_p=0}. This figure shows that for decreasing values of $\beta_{e}$, the distance between the the non-zero saddle and the non-saddle fixed point nearest to it decreases and ultimately both of them disappear from the system. Finally, the system contains only the point of inflexion, i.e., the origin and a non-zero equilibrium point. Consequently, the separatrix corresponding to the solitary structure appears to start and end at the origin enclosing the non-saddle fixed point and we have NPSW after the formation of NPDL at the acoustic speed. Thus, we see that the mechanism of the transition of solitary structures at the acoustic speed is qualitatively same as the transtion process of solitary structures for supersonic waves as reported by Paul \textit{et al.} \cite{paul17pop}. 


\section{\label{sec:Conclusions} Conclusions}

In the present work, we have investigated the nature of existence of different DIA solitary structures at the acoustic speed in a collisionless unmagnetized dusty plasma consisting of negatively charged static dust grains, adiabatic warm ions, Cairns distributed nonthermal electrons and isothermal positrons with the help of existence domains and phase portraits. Although from the paper of Paul \& Bandyopadhyay \cite{paul2016} it has been observed that for supersonic case the same system supports double layers of both polarities and positive potential supersolitons, but at the acoustic speed the system supports PPSWs,  NPSWs and NPDLs only. Again, in the present paper, if we consider $p=0$ then we observe that the system supports PPSWs, NPSWs, NPDLs, NPSWs after the formation of NPDL and negative potential supersolitons at the acoustic speed. These results agree with the results of Das \textit{et al.} \cite{das2011existence}, where they have considered a collisionless unmagnetized three component dusty plasma consisting of negatively charged static dust grains, adiabatic warm ions and Cairns distributed nonthermal electrons to investigate the DIA solitary structures at the acoustic speed.  

For the first time, we have introduced the phase portraits of the dynamical system corresponding to the DIA solitary structures at the acoustic speed. We found the following qualitative differences between the phase portraits of the solitary structures at $M=M_{c}$ and the phase portraits of the solitary structures for $M>M_{c}$ which have been discussed by Paul \textit{et al.} \cite{paul17pop}. (i) For $M>M_{c}$, the origin is always a saddle point and the separatrix corresponding to the solitary structures appears to pass trough the origin, whereas for $M=M_{c}$, the origin is the point of inflexion and the separatrix corresponding to the solitary structures appears to start and end at the origin. (ii) For $M>M_{c}$, the phase portraits of the dynamical system corresponding to DIA  double layers have two saddles and two non-saddle fixed points, but in the case of double layers at the acoustic speed, the system has a point of inflexion at the origin, one non-zero saddle and one non-saddle fixed point. (iii) From the paper of Dubinov and Kolotkov \cite{dubinov12b} and Paul \textit{et al.} \cite{paul17pop}, for the case of supersolitons, we see that there exist at least two separatrices and the separatrix through the origin (saddle point) encloses the other one for $M>M_{c}$. In the case of sonic DIA waves, i.e., for $M=M_{c}$, we have the same definition of the supersolitons, i.e., for supersolitons at the acoustic speed, there are at least two separatrices and the separatrix that appears to start and end 
at the origin (point of inflexion) encloses the other one.

\begin{figure}
\begin{center}
\includegraphics{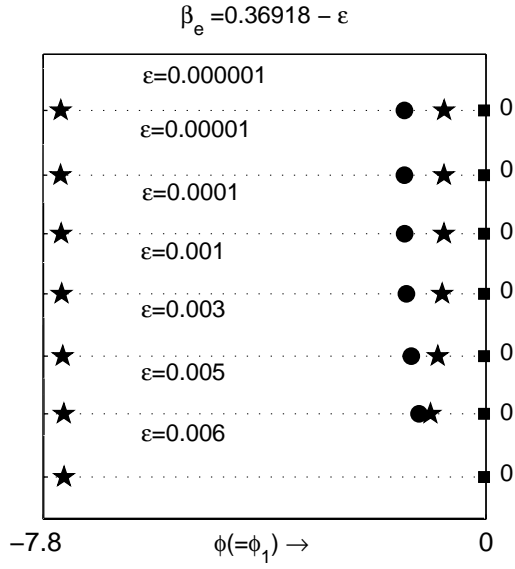}
  \caption{\label{equilibrium_points_p=0} Point of inflexion (small solid cubes), saddle points (small solid circles) and the equilibrium points other than saddle points (small solid stars) for the system (\ref{phase_portraits}) have been drawn on the $\phi$-axis at $M=M_{c}$ for different values of $\beta_{e}$ when $p=0$, $\mu=0.5$ and $\sigma_{ie}=0.9$.}
\end{center}
\end{figure}

With the help of the phase portraits, we have also explained the transition process of the solitary structures at the acoustic speed for $p=0$, viz., soliton $\to$ double layer $\to$ supersoliton $\to$ soliton for decreasing values of $\beta_{e}$ and it is not possible to explain the transition process of solitary structures by considering the existence domains only or simply by drawing the curve $V(\phi)$ against $\phi$. This transition phenomenon at the acoustic speed happens according to the mechanism as described in figure \ref{equilibrium_points_p=0}. Again, the transition mechanism at the acoustic speed is same as that of the supersonic solitary structures as reported by Paul \textit{et al.} \cite{paul17pop}.

From this work, we can conclude that the formation of double layer is also possible at the acoustic speed. Again, according to Alfv{\'e}n \cite{alfven1981}, a double layer consists of two oppositely charged parallel layers resulting in a potential drop in the layer and a vanishing electric field on each side of the layer. Formation of double layers in a plasma system releases an amount of energy which accelerates the charged particles of the system. Above the ionosphere of the Earth acceleration of electrons has been observed in a rather narrow region and the possible cause of such acceleration is the formation of several double layers in that region \cite{alfven1981}. This work is helpful to understand the formation of the double layer at the acoustic speed.

\acknowledgments One of the authors (Ashesh Paul) is thankful to the Department of Science and Technology, Govt. of India, INSPIRE Fellowship Scheme for financial support.

%

\begin{thebibliography}{31}%
\makeatletter
\providecommand \@ifxundefined [1]{%
 \@ifx{#1\undefined}
}%
\providecommand \@ifnum [1]{%
 \ifnum #1\expandafter \@firstoftwo
 \else \expandafter \@secondoftwo
 \fi
}%
\providecommand \@ifx [1]{%
 \ifx #1\expandafter \@firstoftwo
 \else \expandafter \@secondoftwo
 \fi
}%
\providecommand \natexlab [1]{#1}%
\providecommand \enquote  [1]{``#1''}%
\providecommand \bibnamefont  [1]{#1}%
\providecommand \bibfnamefont [1]{#1}%
\providecommand \citenamefont [1]{#1}%
\providecommand \href@noop [0]{\@secondoftwo}%
\providecommand \href [0]{\begingroup \@sanitize@url \@href}%
\providecommand \@href[1]{\@@startlink{#1}\@@href}%
\providecommand \@@href[1]{\endgroup#1\@@endlink}%
\providecommand \@sanitize@url [0]{\catcode `\\12\catcode `\$12\catcode
  `\&12\catcode `\#12\catcode `\^12\catcode `\_12\catcode `\%12\relax}%
\providecommand \@@startlink[1]{}%
\providecommand \@@endlink[0]{}%
\providecommand \url  [0]{\begingroup\@sanitize@url \@url }%
\providecommand \@url [1]{\endgroup\@href {#1}{\urlprefix }}%
\providecommand \urlprefix  [0]{URL }%
\providecommand \Eprint [0]{\href }%
\providecommand \doibase [0]{http://dx.doi.org/}%
\providecommand \selectlanguage [0]{\@gobble}%
\providecommand \bibinfo  [0]{\@secondoftwo}%
\providecommand \bibfield  [0]{\@secondoftwo}%
\providecommand \translation [1]{[#1]}%
\providecommand \BibitemOpen [0]{}%
\providecommand \bibitemStop [0]{}%
\providecommand \bibitemNoStop [0]{.\EOS\space}%
\providecommand \EOS [0]{\spacefactor3000\relax}%
\providecommand \BibitemShut  [1]{\csname bibitem#1\endcsname}%
\let\auto@bib@innerbib\@empty
\bibitem [{\citenamefont {Shukla}\ and\ \citenamefont
  {Marklund}(2004)}]{shukla04}%
  \BibitemOpen
  \bibfield  {author} {\bibinfo {author} {\bibfnamefont {P.~K.}\ \bibnamefont
  {Shukla}}\ and\ \bibinfo {author} {\bibfnamefont {M.}~\bibnamefont
  {Marklund}},\ }\href@noop {} {\bibfield  {journal} {\bibinfo  {journal}
  {Phys. Scr.}\ }\textbf {\bibinfo {volume} {T113}},\ \bibinfo {pages} {36}
  (\bibinfo {year} {2004})}\BibitemShut {NoStop}%
\bibitem [{\citenamefont {Zel'dovich}\ and\ \citenamefont
  {Novikov}(1971)}]{zeldovich1971}%
  \BibitemOpen
  \bibfield  {author} {\bibinfo {author} {\bibfnamefont {I.~B.}\ \bibnamefont
  {Zel'dovich}}\ and\ \bibinfo {author} {\bibfnamefont {I.~D.}\ \bibnamefont
  {Novikov}},\ }\href@noop {} {\emph {\bibinfo {title} {Relativistic
  Astrophysics, 2: The Structure and Evolution of the Universe}}},\
  Vol.~\bibinfo {volume} {2}\ (\bibinfo  {publisher} {University of Chicago
  Press},\ \bibinfo {year} {1971})\BibitemShut {NoStop}%
\bibitem [{\citenamefont {Zurek}(1985)}]{zurek1985}%
  \BibitemOpen
  \bibfield  {author} {\bibinfo {author} {\bibfnamefont {W.}~\bibnamefont
  {Zurek}},\ }\href@noop {} {\bibfield  {journal} {\bibinfo  {journal}
  {Astrophys. J.}\ }\textbf {\bibinfo {volume} {289}},\ \bibinfo {pages} {603}
  (\bibinfo {year} {1985})}\BibitemShut {NoStop}%
\bibitem [{\citenamefont {Higdon}, \citenamefont {Lingenfelter},\ and\
  \citenamefont {Rothschild}(2009)}]{higdon09}%
  \BibitemOpen
  \bibfield  {author} {\bibinfo {author} {\bibfnamefont {J.~C.}\ \bibnamefont
  {Higdon}}, \bibinfo {author} {\bibfnamefont {R.~E.}\ \bibnamefont
  {Lingenfelter}}, \ and\ \bibinfo {author} {\bibfnamefont {R.~E.}\
  \bibnamefont {Rothschild}},\ }\href@noop {} {\bibfield  {journal} {\bibinfo
  {journal} {Astrophys. J.}\ }\textbf {\bibinfo {volume} {698}},\ \bibinfo
  {pages} {350} (\bibinfo {year} {2009})}\BibitemShut {NoStop}%
\bibitem [{\citenamefont {Shukla}(2008)}]{shukla2008}%
  \BibitemOpen
  \bibfield  {author} {\bibinfo {author} {\bibfnamefont {P.~K.}\ \bibnamefont
  {Shukla}},\ }\href@noop {} {\bibfield  {journal} {\bibinfo  {journal} {Phys.
  Scr.}\ }\textbf {\bibinfo {volume} {77}},\ \bibinfo {pages} {068201}
  (\bibinfo {year} {2008})}\BibitemShut {NoStop}%
\bibitem [{\citenamefont {Dubinov}, \citenamefont {Kolotkov},\ and\
  \citenamefont {Sazonkin}(2012)}]{dubinov12}%
  \BibitemOpen
  \bibfield  {author} {\bibinfo {author} {\bibfnamefont {A.~E.}\ \bibnamefont
  {Dubinov}}, \bibinfo {author} {\bibfnamefont {D.~Y.}\ \bibnamefont
  {Kolotkov}}, \ and\ \bibinfo {author} {\bibfnamefont {M.~A.}\ \bibnamefont
  {Sazonkin}},\ }\href@noop {} {\bibfield  {journal} {\bibinfo  {journal}
  {Tech. Phys.}\ }\textbf {\bibinfo {volume} {57}},\ \bibinfo {pages} {585}
  (\bibinfo {year} {2012})}\BibitemShut {NoStop}%
\bibitem [{\citenamefont {Alfv{\'e}n}(1981)}]{alfven1981}%
  \BibitemOpen
  \bibfield  {author} {\bibinfo {author} {\bibfnamefont {H.}~\bibnamefont
  {Alfv{\'e}n}},\ }\href@noop {} {\emph {\bibinfo {title} {Cosmic plasma}}},\
  Vol.~\bibinfo {volume} {82}\ (\bibinfo  {publisher} {D. Reidel Publishing
  Company, Dordrecht:Holland},\ \bibinfo {year} {1981})\BibitemShut {NoStop}%
\bibitem [{\citenamefont {Gusev}\ \emph {et~al.}(2000)\citenamefont {Gusev},
  \citenamefont {Jayanthi}, \citenamefont {Martin}, \citenamefont {Pugacheva},\
  and\ \citenamefont {Spjeldik}}]{gusev2000}%
  \BibitemOpen
  \bibfield  {author} {\bibinfo {author} {\bibfnamefont {A.~A.}\ \bibnamefont
  {Gusev}}, \bibinfo {author} {\bibfnamefont {U.~B.}\ \bibnamefont {Jayanthi}},
  \bibinfo {author} {\bibfnamefont {I.~M.}\ \bibnamefont {Martin}}, \bibinfo
  {author} {\bibfnamefont {G.~I.}\ \bibnamefont {Pugacheva}}, \ and\ \bibinfo
  {author} {\bibfnamefont {W.~N.}\ \bibnamefont {Spjeldik}},\ }\href@noop {}
  {\bibfield  {journal} {\bibinfo  {journal} {Braz. J. Phys.}\ }\textbf
  {\bibinfo {volume} {30}},\ \bibinfo {pages} {590} (\bibinfo {year}
  {2000})}\BibitemShut {NoStop}%
\bibitem [{\citenamefont {Gusev}\ \emph {et~al.}(2001)\citenamefont {Gusev},
  \citenamefont {Jayanthi}, \citenamefont {Martin}, \citenamefont {Pugacheva},\
  and\ \citenamefont {Spjeldvik}}]{gusev2001}%
  \BibitemOpen
  \bibfield  {author} {\bibinfo {author} {\bibfnamefont {A.~A.}\ \bibnamefont
  {Gusev}}, \bibinfo {author} {\bibfnamefont {U.~B.}\ \bibnamefont {Jayanthi}},
  \bibinfo {author} {\bibfnamefont {I.~M.}\ \bibnamefont {Martin}}, \bibinfo
  {author} {\bibfnamefont {G.~I.}\ \bibnamefont {Pugacheva}}, \ and\ \bibinfo
  {author} {\bibfnamefont {W.~N.}\ \bibnamefont {Spjeldvik}},\ }\href@noop {}
  {\bibfield  {journal} {\bibinfo  {journal} {J. Geophys. Res.}\ }\textbf
  {\bibinfo {volume} {106}},\ \bibinfo {pages} {26111} (\bibinfo {year}
  {2001})}\BibitemShut {NoStop}%
\bibitem [{\citenamefont {Merlino}(2006)}]{merlino2006}%
  \BibitemOpen
  \bibfield  {author} {\bibinfo {author} {\bibfnamefont {R.~L.}\ \bibnamefont
  {Merlino}},\ }\href@noop {} {\bibfield  {journal} {\bibinfo  {journal}
  {Plasma Phys. Appl.}\ }\textbf {\bibinfo {volume} {81}},\ \bibinfo {pages}
  {73} (\bibinfo {year} {2006})}\BibitemShut {NoStop}%
\bibitem [{\citenamefont {Hor{\'a}nyi}\ \emph {et~al.}(2004)\citenamefont
  {Hor{\'a}nyi}, \citenamefont {Hartquist}, \citenamefont {Havnes},
  \citenamefont {Mendis},\ and\ \citenamefont {Morfill}}]{horanyi2004}%
  \BibitemOpen
  \bibfield  {author} {\bibinfo {author} {\bibfnamefont {M.}~\bibnamefont
  {Hor{\'a}nyi}}, \bibinfo {author} {\bibfnamefont {T.}~\bibnamefont
  {Hartquist}}, \bibinfo {author} {\bibfnamefont {O.}~\bibnamefont {Havnes}},
  \bibinfo {author} {\bibfnamefont {D.}~\bibnamefont {Mendis}}, \ and\ \bibinfo
  {author} {\bibfnamefont {G.}~\bibnamefont {Morfill}},\ }\href@noop {}
  {\bibfield  {journal} {\bibinfo  {journal} {Rev. Geophys.}\ }\textbf
  {\bibinfo {volume} {42}},\ \bibinfo {pages} {RG4002} (\bibinfo {year}
  {2004})}\BibitemShut {NoStop}%
\bibitem [{\citenamefont {Ghosh}\ and\ \citenamefont
  {Bharuthram}(2008)}]{ghosh08}%
  \BibitemOpen
  \bibfield  {author} {\bibinfo {author} {\bibfnamefont {S.}~\bibnamefont
  {Ghosh}}\ and\ \bibinfo {author} {\bibfnamefont {R.}~\bibnamefont
  {Bharuthram}},\ }\href@noop {} {\bibfield  {journal} {\bibinfo  {journal}
  {Astrophys. Space Sci.}\ }\textbf {\bibinfo {volume} {314}},\ \bibinfo
  {pages} {121} (\bibinfo {year} {2008})}\BibitemShut {NoStop}%
\bibitem [{\citenamefont {El-Tantawy}, \citenamefont {El-Bedwehy},\ and\
  \citenamefont {Moslem}(2011)}]{el-tantawy11a}%
  \BibitemOpen
  \bibfield  {author} {\bibinfo {author} {\bibfnamefont {S.~A.}\ \bibnamefont
  {El-Tantawy}}, \bibinfo {author} {\bibfnamefont {N.~A.}\ \bibnamefont
  {El-Bedwehy}}, \ and\ \bibinfo {author} {\bibfnamefont {W.~M.}\ \bibnamefont
  {Moslem}},\ }\href@noop {} {\bibfield  {journal} {\bibinfo  {journal} {Phys.
  Plasmas}\ }\textbf {\bibinfo {volume} {18}},\ \bibinfo {pages} {052113}
  (\bibinfo {year} {2011})}\BibitemShut {NoStop}%
\bibitem [{\citenamefont {El-Tantawy}\ and\ \citenamefont
  {Moslem}(2011)}]{el-tantawy11b}%
  \BibitemOpen
  \bibfield  {author} {\bibinfo {author} {\bibfnamefont {S.~A.}\ \bibnamefont
  {El-Tantawy}}\ and\ \bibinfo {author} {\bibfnamefont {W.~M.}\ \bibnamefont
  {Moslem}},\ }\href@noop {} {\bibfield  {journal} {\bibinfo  {journal} {Phys.
  Plasmas}\ }\textbf {\bibinfo {volume} {18}},\ \bibinfo {pages} {112105}
  (\bibinfo {year} {2011})}\BibitemShut {NoStop}%
\bibitem [{\citenamefont {Saini}, \citenamefont {Chahal},\ and\ \citenamefont
  {Bains}(2013)}]{saini13}%
  \BibitemOpen
  \bibfield  {author} {\bibinfo {author} {\bibfnamefont {N.~S.}\ \bibnamefont
  {Saini}}, \bibinfo {author} {\bibfnamefont {B.~S.}\ \bibnamefont {Chahal}}, \
  and\ \bibinfo {author} {\bibfnamefont {A.~S.}\ \bibnamefont {Bains}},\
  }\href@noop {} {\bibfield  {journal} {\bibinfo  {journal} {Astrophys. Space
  Sci.}\ }\textbf {\bibinfo {volume} {347}},\ \bibinfo {pages} {129} (\bibinfo
  {year} {2013})}\BibitemShut {NoStop}%
\bibitem [{\citenamefont {Banerjee}\ and\ \citenamefont
  {Maitra}(2016)}]{banerjee16}%
  \BibitemOpen
  \bibfield  {author} {\bibinfo {author} {\bibfnamefont {G.}~\bibnamefont
  {Banerjee}}\ and\ \bibinfo {author} {\bibfnamefont {S.}~\bibnamefont
  {Maitra}},\ }\href@noop {} {\bibfield  {journal} {\bibinfo  {journal} {Phys.
  Plasmas}\ }\textbf {\bibinfo {volume} {23}},\ \bibinfo {pages} {123701}
  (\bibinfo {year} {2016})}\BibitemShut {NoStop}%
\bibitem [{\citenamefont {Paul}, \citenamefont {Das},\ and\ \citenamefont
  {Bandyopadhyay}(2017)}]{paul17ppr}%
  \BibitemOpen
  \bibfield  {author} {\bibinfo {author} {\bibfnamefont {A.}~\bibnamefont
  {Paul}}, \bibinfo {author} {\bibfnamefont {A.}~\bibnamefont {Das}}, \ and\
  \bibinfo {author} {\bibfnamefont {A.}~\bibnamefont {Bandyopadhyay}},\
  }\href@noop {} {\bibfield  {journal} {\bibinfo  {journal} {Plasma Phys.
  Rep.}\ }\textbf {\bibinfo {volume} {43}},\ \bibinfo {pages} {218} (\bibinfo
  {year} {2017})}\BibitemShut {NoStop}%
\bibitem [{\citenamefont {Paul}\ and\ \citenamefont
  {Bandyopadhyay}(2016)}]{paul2016}%
  \BibitemOpen
  \bibfield  {author} {\bibinfo {author} {\bibfnamefont {A.}~\bibnamefont
  {Paul}}\ and\ \bibinfo {author} {\bibfnamefont {A.}~\bibnamefont
  {Bandyopadhyay}},\ }\href@noop {} {\bibfield  {journal} {\bibinfo  {journal}
  {Astrophys. Space Sci.}\ }\textbf {\bibinfo {volume} {361}},\ \bibinfo
  {pages} {172} (\bibinfo {year} {2016})}\BibitemShut {NoStop}%
\bibitem [{\citenamefont {Cairns}\ \emph {et~al.}(1995)\citenamefont {Cairns},
  \citenamefont {Mamun}, \citenamefont {Bingham}, \citenamefont {Dendy},
  \citenamefont {Bostr{\"o}m}, \citenamefont {Shukla},\ and\ \citenamefont
  {Nairn}}]{cairns95}%
  \BibitemOpen
  \bibfield  {author} {\bibinfo {author} {\bibfnamefont {R.~A.}\ \bibnamefont
  {Cairns}}, \bibinfo {author} {\bibfnamefont {A.~A.}\ \bibnamefont {Mamun}},
  \bibinfo {author} {\bibfnamefont {R.}~\bibnamefont {Bingham}}, \bibinfo
  {author} {\bibfnamefont {R.~O.}\ \bibnamefont {Dendy}}, \bibinfo {author}
  {\bibfnamefont {R.}~\bibnamefont {Bostr{\"o}m}}, \bibinfo {author}
  {\bibfnamefont {P.~K.}\ \bibnamefont {Shukla}}, \ and\ \bibinfo {author}
  {\bibfnamefont {C.~M.~C.}\ \bibnamefont {Nairn}},\ }\href@noop {} {\bibfield
  {journal} {\bibinfo  {journal} {Geophys. Res. Lett.}\ }\textbf {\bibinfo
  {volume} {22}},\ \bibinfo {pages} {2709} (\bibinfo {year}
  {1995})}\BibitemShut {NoStop}%
\bibitem [{\citenamefont {Paul}, \citenamefont {Bandyopadhyay},\ and\
  \citenamefont {Das}(2017)}]{paul17pop}%
  \BibitemOpen
  \bibfield  {author} {\bibinfo {author} {\bibfnamefont {A.}~\bibnamefont
  {Paul}}, \bibinfo {author} {\bibfnamefont {A.}~\bibnamefont {Bandyopadhyay}},
  \ and\ \bibinfo {author} {\bibfnamefont {K.~P.}\ \bibnamefont {Das}},\
  }\href@noop {} {\bibfield  {journal} {\bibinfo  {journal} {Phys. Plasmas}\
  }\textbf {\bibinfo {volume} {24}},\ \bibinfo {pages} {013707} (\bibinfo
  {year} {2017})}\BibitemShut {NoStop}%
\bibitem [{\citenamefont {Baluku}\ \emph {et~al.}(2010)\citenamefont {Baluku},
  \citenamefont {Hellberg}, \citenamefont {Kourakis},\ and\ \citenamefont
  {Saini}}]{baluku10}%
  \BibitemOpen
  \bibfield  {author} {\bibinfo {author} {\bibfnamefont {T.~K.}\ \bibnamefont
  {Baluku}}, \bibinfo {author} {\bibfnamefont {M.~A.}\ \bibnamefont
  {Hellberg}}, \bibinfo {author} {\bibfnamefont {I.}~\bibnamefont {Kourakis}},
  \ and\ \bibinfo {author} {\bibfnamefont {N.~S.}\ \bibnamefont {Saini}},\
  }\href@noop {} {\bibfield  {journal} {\bibinfo  {journal} {Phys. Plasmas}\
  }\textbf {\bibinfo {volume} {17}},\ \bibinfo {pages} {053702} (\bibinfo
  {year} {2010})}\BibitemShut {NoStop}%
\bibitem [{\citenamefont {Baluku}, \citenamefont {Hellberg},\ and\
  \citenamefont {Verheest}(2010)}]{baluku10a}%
  \BibitemOpen
  \bibfield  {author} {\bibinfo {author} {\bibfnamefont {T.~K.}\ \bibnamefont
  {Baluku}}, \bibinfo {author} {\bibfnamefont {M.~A.}\ \bibnamefont
  {Hellberg}}, \ and\ \bibinfo {author} {\bibfnamefont {F.}~\bibnamefont
  {Verheest}},\ }\href@noop {} {\bibfield  {journal} {\bibinfo  {journal}
  {Europhys. Lett.}\ }\textbf {\bibinfo {volume} {91}},\ \bibinfo {pages}
  {15001} (\bibinfo {year} {2010})}\BibitemShut {NoStop}%
\bibitem [{\citenamefont {Verheest}\ and\ \citenamefont
  {Hellberg}(2010)}]{verheest10}%
  \BibitemOpen
  \bibfield  {author} {\bibinfo {author} {\bibfnamefont {F.}~\bibnamefont
  {Verheest}}\ and\ \bibinfo {author} {\bibfnamefont {M.~A.}\ \bibnamefont
  {Hellberg}},\ }\href@noop {} {\bibfield  {journal} {\bibinfo  {journal}
  {Phys. Plasmas}\ }\textbf {\bibinfo {volume} {17}},\ \bibinfo {pages}
  {023701} (\bibinfo {year} {2010})}\BibitemShut {NoStop}%
\bibitem [{\citenamefont {Das}, \citenamefont {Bandyopadhyay},\ and\
  \citenamefont {Das}(2012{\natexlab{a}})}]{das12mc}%
  \BibitemOpen
  \bibfield  {author} {\bibinfo {author} {\bibfnamefont {A.}~\bibnamefont
  {Das}}, \bibinfo {author} {\bibfnamefont {A.}~\bibnamefont {Bandyopadhyay}},
  \ and\ \bibinfo {author} {\bibfnamefont {K.~P.}\ \bibnamefont {Das}},\
  }\href@noop {} {\bibfield  {journal} {\bibinfo  {journal} {J. Plasma Phys.}\
  }\textbf {\bibinfo {volume} {78}},\ \bibinfo {pages} {565} (\bibinfo {year}
  {2012}{\natexlab{a}})}\BibitemShut {NoStop}%
\bibitem [{\citenamefont {Das}, \citenamefont {Bandyopadhyay},\ and\
  \citenamefont {Das}(2011)}]{das2011existence}%
  \BibitemOpen
  \bibfield  {author} {\bibinfo {author} {\bibfnamefont {A.}~\bibnamefont
  {Das}}, \bibinfo {author} {\bibfnamefont {A.}~\bibnamefont {Bandyopadhyay}},
  \ and\ \bibinfo {author} {\bibfnamefont {K.}~\bibnamefont {Das}},\ }\bibfield
   {title} {\enquote {\bibinfo {title} {Existence of dust ion acoustic solitary
  wave and double layer solution at m= mc},}\ }\href@noop {} {\bibfield
  {journal} {\bibinfo  {journal} {arXiv preprint arXiv:1110.5307}\ } (\bibinfo
  {year} {2011})}\BibitemShut {NoStop}%
\bibitem [{\citenamefont {Verheest}\ and\ \citenamefont
  {Hellberg}(2015)}]{verheest2015}%
  \BibitemOpen
  \bibfield  {author} {\bibinfo {author} {\bibfnamefont {F.}~\bibnamefont
  {Verheest}}\ and\ \bibinfo {author} {\bibfnamefont {M.~A.}\ \bibnamefont
  {Hellberg}},\ }\href@noop {} {\bibfield  {journal} {\bibinfo  {journal}
  {Phys. Plasmas}\ }\textbf {\bibinfo {volume} {22}},\ \bibinfo {pages}
  {012301} (\bibinfo {year} {2015})}\BibitemShut {NoStop}%
\bibitem [{\citenamefont {Verheest}\ and\ \citenamefont
  {Pillay}(2008)}]{verheest08}%
  \BibitemOpen
  \bibfield  {author} {\bibinfo {author} {\bibfnamefont {F.}~\bibnamefont
  {Verheest}}\ and\ \bibinfo {author} {\bibfnamefont {S.~R.}\ \bibnamefont
  {Pillay}},\ }\href@noop {} {\bibfield  {journal} {\bibinfo  {journal} {Phys.
  Plasmas}\ }\textbf {\bibinfo {volume} {15}},\ \bibinfo {pages} {013703}
  (\bibinfo {year} {2008})}\BibitemShut {NoStop}%
\bibitem [{\citenamefont {Sagdeev}(1966)}]{sagdeev66}%
  \BibitemOpen
  \bibfield  {author} {\bibinfo {author} {\bibfnamefont {R.~Z.}\ \bibnamefont
  {Sagdeev}},\ }\href@noop {} {\emph {\bibinfo {title} {Reviews of Plasma
  Physics Vol-4 (ed. M. A. Leontovich)}}}\ (\bibinfo  {publisher} {New York,
  NY: Consultant Bureau},\ \bibinfo {year} {1966})\BibitemShut {NoStop}%
\bibitem [{\citenamefont {Das}, \citenamefont {Bandyopadhyay},\ and\
  \citenamefont {Das}(2009)}]{das09}%
  \BibitemOpen
  \bibfield  {author} {\bibinfo {author} {\bibfnamefont {A.}~\bibnamefont
  {Das}}, \bibinfo {author} {\bibfnamefont {A.}~\bibnamefont {Bandyopadhyay}},
  \ and\ \bibinfo {author} {\bibfnamefont {K.~P.}\ \bibnamefont {Das}},\
  }\href@noop {} {\bibfield  {journal} {\bibinfo  {journal} {Phys. Plasmas}\
  }\textbf {\bibinfo {volume} {16}},\ \bibinfo {pages} {073703} (\bibinfo
  {year} {2009})}\BibitemShut {NoStop}%
\bibitem [{\citenamefont {Das}, \citenamefont {Bandyopadhyay},\ and\
  \citenamefont {Das}(2012{\natexlab{b}})}]{das12}%
  \BibitemOpen
  \bibfield  {author} {\bibinfo {author} {\bibfnamefont {A.}~\bibnamefont
  {Das}}, \bibinfo {author} {\bibfnamefont {A.}~\bibnamefont {Bandyopadhyay}},
  \ and\ \bibinfo {author} {\bibfnamefont {K.~P.}\ \bibnamefont {Das}},\
  }\href@noop {} {\bibfield  {journal} {\bibinfo  {journal} {J. Plasma Phys.}\
  }\textbf {\bibinfo {volume} {78}},\ \bibinfo {pages} {149} (\bibinfo {year}
  {2012}{\natexlab{b}})}\BibitemShut {NoStop}%
\bibitem [{\citenamefont {Dubinov}\ and\ \citenamefont
  {Kolotkov}(2012)}]{dubinov12b}%
  \BibitemOpen
  \bibfield  {author} {\bibinfo {author} {\bibfnamefont {A.~E.}\ \bibnamefont
  {Dubinov}}\ and\ \bibinfo {author} {\bibfnamefont {D.~Y.}\ \bibnamefont
  {Kolotkov}},\ }\href@noop {} {\bibfield  {journal} {\bibinfo  {journal}
  {Plasma Phys. Rep.}\ }\textbf {\bibinfo {volume} {38}},\ \bibinfo {pages}
  {909} (\bibinfo {year} {2012})}\BibitemShut {NoStop}%
\end{thebibliography}
%

\providecommand{\noopsort}[1]{}\providecommand{\singleletter}[1]{#1}%

\newpage


\end{document}